\documentclass[twocolumn]{aastex631}

\usepackage[normalem]{ulem}
\usepackage{xcolor}
\usepackage{amssymb}
\usepackage{amsmath}
\usepackage{graphicx}
\usepackage{float}
\usepackage{physics}
\let\tablenum\relax
\usepackage{siunitx}

\DeclareSIUnit{\elecd}{\meter\squared\per\second}

\makeatletter
\newcommand\newtag[2]{#1\def\@currentlabel{#1}\label{#2}}
\makeatother

\newcommand{\B}{{\bf B}}
\newcommand{\E}{{\bf E}}
\newcommand{\J}{{\bf J}}
\newcommand{\F}{{\bf F}}

\newcommand{\vi}{{\bf v}}

\newcommand{\Bt}{{\tilde B}}

\newcommand{\vt}{{\tilde v}}

\newcommand{\ki}{{\bf k}}
\newcommand{\ri}{{\bf r}}
\newcommand{\e}{{\bf e}}

\newcommand{\x}{{\boldsymbol \xi}}

\newcommand{\wb}{\bar \omega}

\begin{document}

\title{A non-local magneto-curvature instability in a differentially rotating disk}  

\author{Fatima Ebrahimi}
\affiliation{Princeton Plasma Physics Laboratory, Princeton University, Princeton, New Jersey 08543, USA}
\affiliation{Department of Astrophysical Sciences, Princeton University, Princeton, New Jersey 08544, USA}
\author{Matthew Pharr}
\affiliation{Princeton Plasma Physics Laboratory, Princeton University, Princeton, New Jersey 08543, USA}
\affiliation{Applied Physics and Applied Mathematics, Columbia University, New York, New York, 10027, USA}


\begin{abstract}
A global mode is shown to be unstable to non-axisymmetric perturbations in a differentially rotating Keplerian disk containing either vertical or azimuthal magnetic fields. In an unstratified cylindrical disk model, using both global eigenvalue stability analysis and linear global initial-value simulations, it is demonstrated that this instability dominates at strong magnetic field where local standard MRI becomes stable. Unlike the standard MRI mode, which is concentrated in the high flow shear region, these distinct global modes (with low azimuthal mode numbers) are extended in the global domain and are Alfv\'en continuum driven unstable modes. As its mode structure and relative dominance over MRI is inherently determined by the global spatial curvature as well as the flow shear in the presence of magnetic field, we call it the magneto-curvature (magneto-spatial-curvature) instability. Consistent with the linear analysis, as the field strength is increased in the nonlinear simulations, a transition from a MRI-driven turbulence to a state dominated by global non-axisymmetric modes is obtained. This global instability could therefore be a source of nonlinear transport in accretion disks at higher magnetic field than predicted by local models. 
\end{abstract}

\section{Introduction}

Search for instabilities to explain the accretion rates in flow-dominated astrophysical settings with massive central objects has led to great interest in the magnetorotational instability (MRI)~\citep{velikhov1959,chandrasekhar60,Balbus1991}. It was
recognized that weak magnetic fields could trigger an MHD instability in astrophysical disks, which could potentially cause turbulent-enhanced viscosity to account for the rapid angular momentum transport~\citep{Balbus1998}. Analytical and numerical studies using both local shearing box~\citep{hawley05,stone96,Fromang_2007, fromang2,lesur07,pessah07,bodo08} and global ~\citep{machida00,hawley00,goodman02} simulations have demonstrated a 3-D MHD turbulence state, as well as dynamo sustainment~\citep{brandenburg_1995,Lesur_2008,Rincon_2007, Ebrahimi_2009} by the MRI. Most of these studies are either limited to the shearing box approximations or are too complex in full global models~\citep{beckwith2008} to clarify the underlying physical nature of the global instabilities. In this paper, via a first-principle, intermediate 
approach, i.e. in a real global domain with spatial curvature,  we examine the onset of global instabilities in rotating systems,  as well as their nonlinear evolution. We uncover globally extended MHD flow-driven modes, which are absent in the local simulations. 

Global axisymmetric MRI (radially uniform channel modes in the shearing box approximation), and non-axisymmetric modes are the primary modes in a turbulent-driven MRI state. In particular non-axisymmetric  MRI modes are rather locally concentrated~\citep{tajima05,Ogilvie_1996} in the region of flow shear, while exponentially growing axisymmetric modes \citep{GoodmanXu1994} could be the primary driver. Localized  non-axisymmetric MRI mode structures were found to be confined between the Alfvénic resonant points (where the magnitude of the Doppler-shifted wave frequency is equal to Alfvén frequency ~\citep{tajima05}. In cylindrical shear flows~\citep{Ogilvie_1996} and in the compressible limit~\citep{goedbloed2022}, it was shown that forward and backward overlapping of these Alfvén continua results in discrete localized non-axisymmetric modes at large axial and azimuthal mode numbers. 
The question arises whether in a real domain with spatial curvature, global non-axisymmetric modes with real frequencies 
can persist. Here, we find distinct global (i.e. large-scale) modes, which are Alfv\'en-continuum-driven unstable modes due to global differential rotation and magnetic curvature.

In this paper, using a hierarchy of models, we examine the global stability of differentially rotating disk systems threaded with either vertical or azimuthal magnetic fields. We find that due to the spatial curvature,  globally extended non-axisymmetric modes (with low azimuthal number) are unstable at a wider range of $V_A/V_0$, beyond the MRI stability limit. For a differentially rotating Keplerian disk,  unlike the standard MRI mode, which is concentrated in the high shear region, it is shown that these global modes are extended in the domain (and with two Alfvénic points). We find that due to the inherent presence of global flow shear and curvature,  Alfvén continua can provide the free energy for a distinct global mode (even at low axial and azimuthal mode number,  $k_z$ and m). These modes remain unstable at stronger magnetic field where standard MRI modes are stable. Furthermore, consistent with the linear global analysis, as $V_A/V_0$ is increased,  a transition from a MRI-driven turbulence to a much more laminar state, dominated by a global non-axisymmetric mode, is also observed in the nonlinear global simulations.

We start by introducing our models in Section 1.1. In flow-dominated  astrophysical systems,  local approximations, due to their simplicity,  have been the most commonly used analysis. Here we also first employ  the local stability analysis with general magnetic field in section \ref{sec:local} and discuss the effect of magnetic and spatial curvature. In section \ref{sec:global}, we present the global sets of equations,  and the resulting ordinary differential equation (ODE) with both vertical and azimuthal fields. Unlike general global pressure or current driven instabilities, where an integrated linearized self-adjoint force operator~\cite{freidberg_rwm} is used to realise the free energy in the flow-less MHD, a modified energy principle is required~\citep{hameiri1990} with flows (due to non-self-adjoint  property of the MHD force operator as per~\cite{frieman_rotenberg}). 
Using the modified energy principle, we examine the additional energy terms in the differentially rotating system in section \ref{sec:global}. We then present the eigenvalue global solutions for vertical and azimuthal fields. The initial-value nonlinear extended MHD code NIMROD, is used to verify the eigenmode solutions in the linear limit. In section \ref{sec:nonlin}, direct nonlinear simulations are presented. We summarize our new findings in section \ref{sec:summary}. 

\subsection{Hierarchy of models}
Here, we employ a hierarchy of physics models and computational techniques for a Keplerian disk, including local and global linear stability analysis, as well as nonlinear simulations. We first examine the local stability analysis using WKB approximation, and obtain a general MHD dispersion relation in a cylinder. Second, linear global stability of Keplerian flows is studied in the incompressible regime by obtaining an ordinary differential equation (ODE)  of the system. Numerical solutions of the global study are presented through both solving the ODE using the shooting method and by performing direct linear simulations using the NIMROD code. For completeness and verification, the linear results from shooting and NIMROD simulations are compared in the case with the pure azimuthal magnetic field. We finally present nonlinear results, by performing global nonlinear simulations using NIMROD code. 

For our direct numerical simulations of a Keplerian disk, we employ the NIMROD code. NIMROD (Non-Ideal Magnetohydrodynamics 
with Rotation, an Open Discussion project) code~\citep{SOVINEC2004355} solves the 3-D, nonlinear, time-dependent, compressible, extended MHD equations. The discrete form of the equations
in NIMROD uses high-order finite elements to represent the poloidal plane (r-z) and 
is pseudo-spectral with FFTs for the periodic direction ($\phi$). The basis functions of the finite elements are polynomials. The finite element spatial discretization allows arbitrary poloidal cross sections and flexibility in the geometry. The full extended MHD model in 
NIMROD includes resistive, two-fluid 
and Finite Larmor Radius (FLR) effects~\citep{flr,Ebrahimi_2011_hall}. In this paper, we model a differentially rotating system with an initial magnetic field in an unstratified Keplerian cylinder ($r$,$z$,$\phi$), where a finite element discretization in the poloidal 
plane ($r$-$z$) and a spectral representation in the azimuthal ($\phi$) direction is employed. We use a similar set-up as described in \citet{Ebrahimi_2011_hall} in both linear and nonlinear MHD simulations. 
As NIMROD is an initial value code, in the linear NIMROD computations, the initial conditions consist of an equilibrium $\langle f(r)\rangle$ plus 
a mode with a single \textit{m} perturbation $\tilde f_{m}(r,z,0) \mathrm \exp( im\phi)$. 
Only the mode with a specific \textit{m} is then evolved; in particular,  
the equilibrium ($\langle f\rangle$) is \textit{not} 
evolved, and remains fixed in time. In the fully nonlinear computations, all modes (axisymmetric \textit{m}=0 and non-axisymmetric $m\neq 0$) 
are initialized 
with small random amplitude and are evolved in time, 
including the full nonlinear terms. An impenetrable radial boundary condition is used, while the azimuthal and vertical (z) directions are periodic. Starting with a current free equilibrium (by applying a uniform magnetic field or a pure azimuthal field), the disk rotate azimuthally with a mean Keplerian flow $\langle V_{\phi}(r)\rangle \propto r^{-1/2} $. The set of NIMROD simulations with the relevant dimensionless parameters used in this paper, namely the magnetic Reynolds, Lundquist number, and magnetic Prandtl numbers, are given in Table \ref{tab:simulations}.
\begin{figure}
        (a) \hspace{40mm} (b)\\
        \includegraphics[width=0.49\columnwidth]{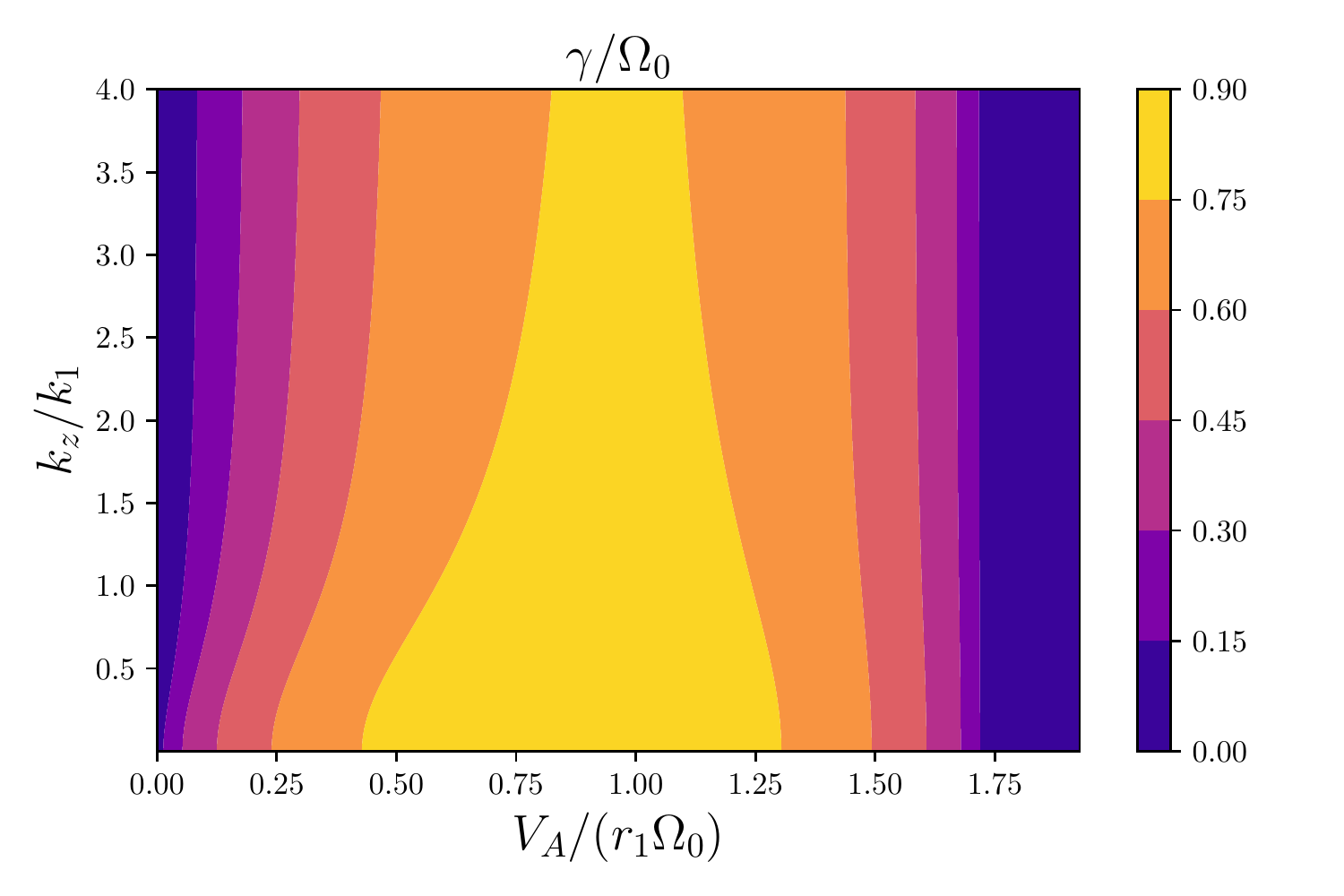}
        \includegraphics[width=0.49\columnwidth]{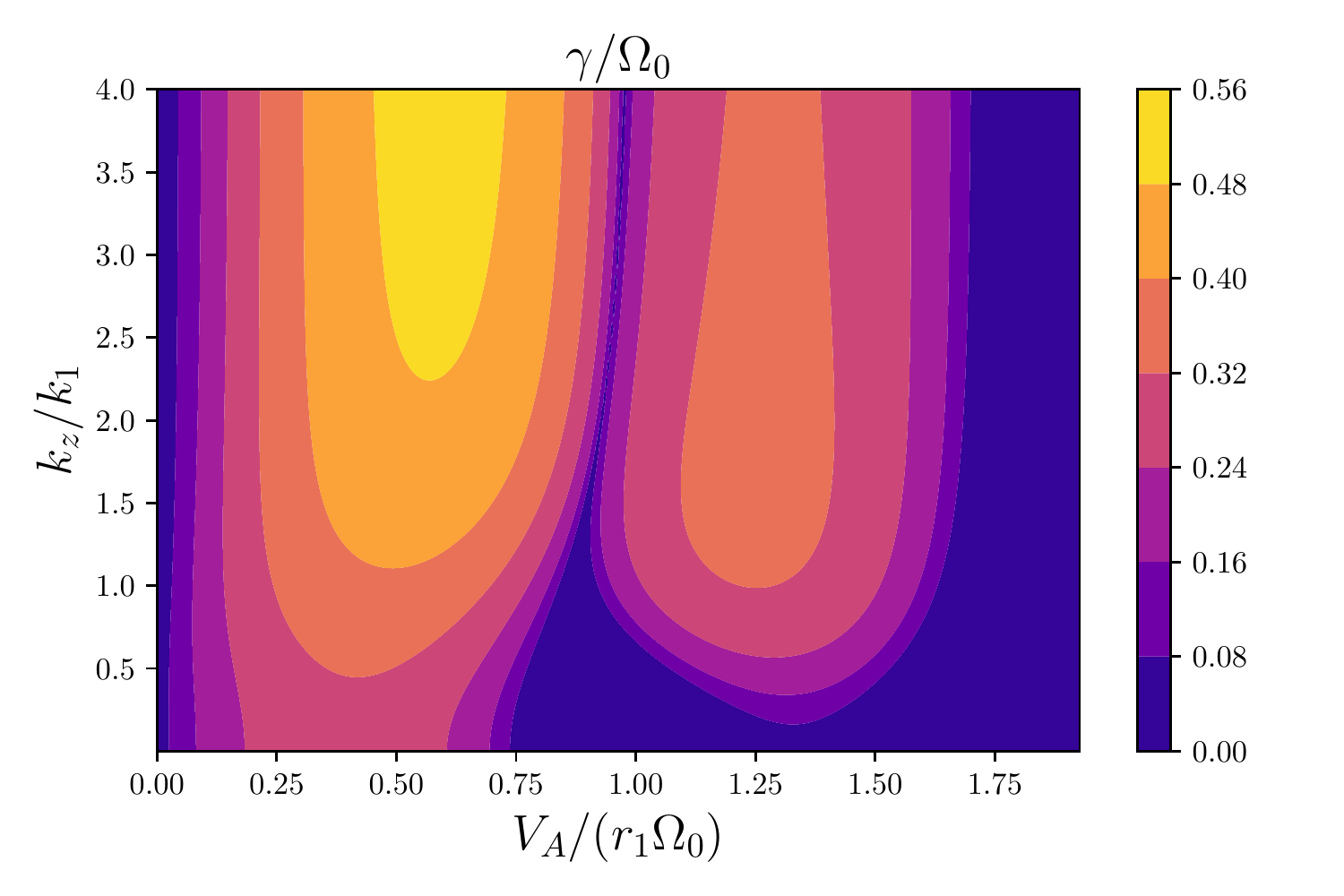}
        (c) \hspace{40mm} (d)\\
        \includegraphics[width=0.49\columnwidth]{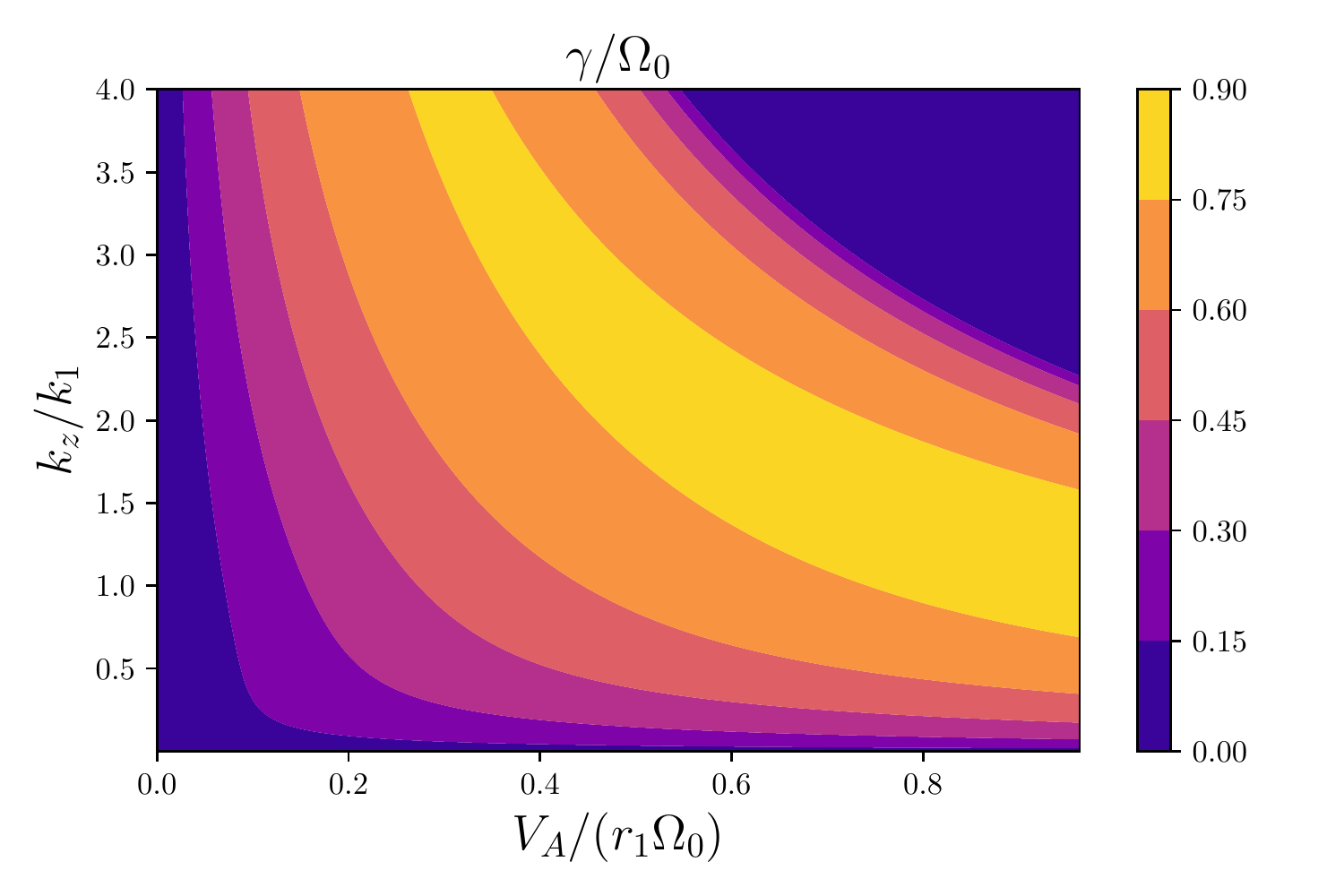}
        \includegraphics[width=0.49\columnwidth]{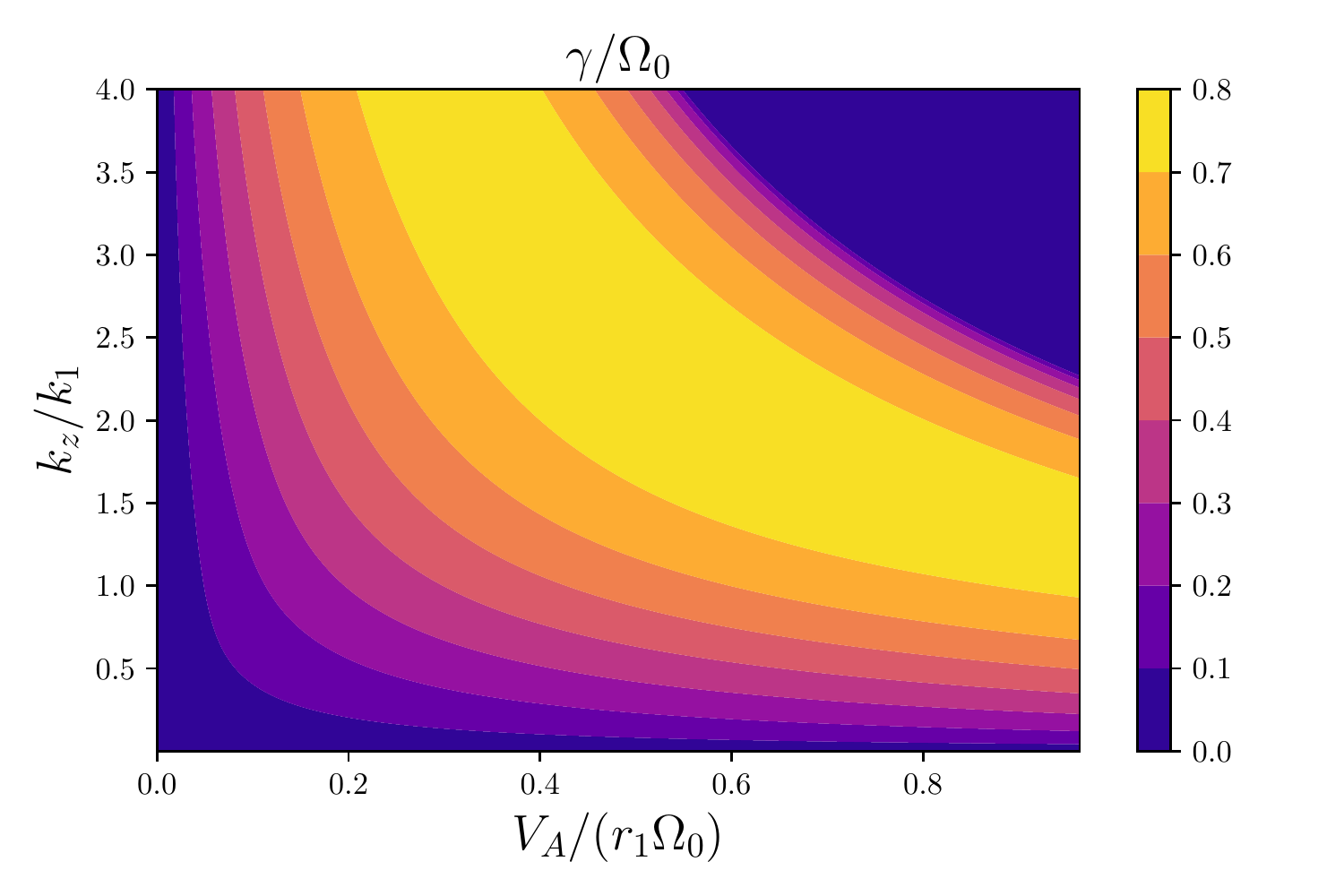}
    \caption{Contour plots of normalized WKB growth rates from  (eqs \ref{eq:mat1} - \ref{eq:mat4}) in the  $k_z/k_1$  and $V_A/r_1 \Omega_1$ plane for $k_r = 0$ (a) without $\delta_c$  (in the cartesian limit) (b) with $\delta_c=1$ with pure toroidal magnetic field. (c) and (d) are equivalent instead with pure vertical magnetic field.}
    \label{fig:disp_rel_2d}
\end{figure}

%

\section{Local stability Analysis}\label{sec:local}

\begin{figure}
    \centering
    \includegraphics[width=0.9\columnwidth]{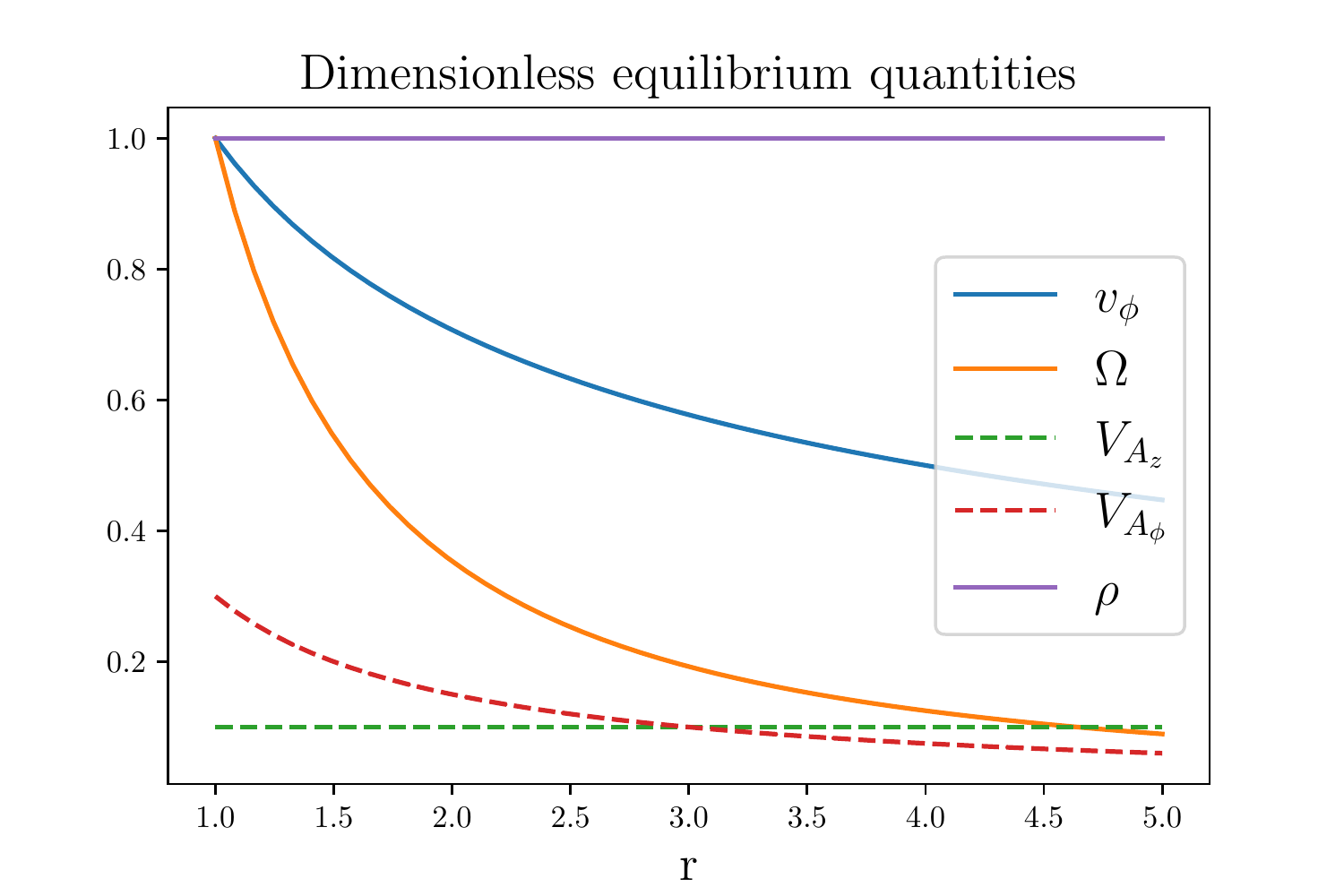}
    \caption{Equilibrium quantities for an unstratified, currentless Keplerian disk. As magnetic field may vary, $V_A$ was arbitrarily chosen such that ${V_A}_z = 0.1 r_1\Omega_0$ and ${{{V_A}_\phi}_0} = 0.3r_1\Omega_0$ for demonstration purposes.}
    \label{fig:equil}
\end{figure}

We begin with the momentum and induction equations:

\begin{align}
    \label{eq:induction}
    \frac{\partial \B}{\partial t} &= \nabla \times\left( \vi \times \B \right) - \eta\nabla\times\J\\
    \label{eq:momentum}
    \rho \left( \frac{\partial}{\partial t} + \vi \cdot \nabla \right)\vi &= \frac{\left(\B\cdot\nabla\right)\B}{\mu_0} - \nabla \left(p + \frac{B^2}{2\mu_0}\right)+\rho\nu\nabla^2\vi
\end{align}

where $\vi$, $\B$, $\J$, $p$, $\eta$, and $\nu$ are the fluid velocity, magnetic field, current, pressure, and the electric and kinetic diffusivities, respectively. We consider perturbed quantities in the form of $ \x(\ri,t) = \x_0 \exp\left\{i(\ki\cdot\ri- \omega t)\right\}$, defining $\omega = \omega_r + i\gamma$ and $\ki\cdot\ri = -i k_r r +m \phi + k_z z$. We have chosen this construction for $\ki$ so that for real $k_r$, the solutions will be evanescent in the radially outward direction and to eliminate what would otherwise be complex terms in Eq.~(\ref{eq:c3})-(\ref{eq:c0}). We assume an equilibrium with $\B_0 = B_\phi(r)\,\hat\phi + B_z\,\hat z$ and $\vi_0 = v_\phi(r)$, where the radially dependent quantities are constrained by an assumption of a currentless magnetic field and a Keplerian flow profile, as seen in Fig.~\ref{fig:equil}. Additionally, we chose to assume a constant radial density for the sake of simplicity in investigating the global behaviors of Alfv\'enic Resonance Instabilities; much complexity could be added by considering different density profiles, so we leave it a topic for future study.

\begin{equation}
   \label{eq:mat1}
    i\wb \Bt_r + 0 \Bt_\phi + \left(i F \right) \vt_r + 0\vt_\phi = 0\\
    \end{equation}
    \begin{multline}
    \left(\frac{\partial \Omega}{\partial \ln r}\right)\Bt_r +i\wb \Bt_\phi + \left(\frac{2B_\phi}{r}\delta_c\right)\vt_r + \left(i F \right) \vt_\phi = 0\\
    \end{multline}
    \begin{multline}
    \frac{i F}{\mu_0 \rho}\left(1-\frac{k_r \bar{k_r}}{k_z^2}\right) \Bt_r + \frac{1}{\mu_0 \rho}\left(\frac{F k_\phi k_r}{k_z^2} - \frac{1}{r}\delta_c B_\phi \right)\Bt_\phi \\+i\wb \left(1-\frac{k_r \bar{k_r}}{k_z^2}\right)\vt_r + \left(2\Omega +\wb \frac{k_rk_\phi}{k_z^2}\right)\vt_\phi=0 \\
    \end{multline}
    \begin{multline}
    \label{eq:mat4}
    \frac{F}{\mu_0 \rho}\left(\frac{k_\phi \bar{k_r}}{k_z^2}\right)\Bt_r + \frac{iF}{\mu_0\rho}\left(1+\frac{k_\phi^2}{k_z^2}\right)\Bt_\phi \\+ \left(\wb\frac{k_\phi\bar{k_r}}{k_z^2}-\frac{\kappa^2}{2\Omega}\right)\vt_r + i\wb\left(\frac{k_\phi^2}{k_z^2}+1\right) \vt_\phi = 0\\
\end{multline}

where $k_\phi \equiv m/r$, $\bar{k_r} = (k_r+\delta_c/r)$, $F\equiv\ki\cdot \B = k_\phi B_\phi(r) + k_z B_z$, and $\kappa$ is the epicyclic frequency, where $\kappa^2 \equiv 4\Omega^2 + \frac{\partial \Omega^2}{\partial \ln r}$. Notably, for quasi-Keplerian disks being examined in this paper, $\kappa = \Omega$. Using the curvilinear definition of $\nabla$ in all derivatives gives rise to the curvature term for the toroidal magnetic field, $2 B_\phi/r$, accompanied by $\delta_c$. The notation $\delta_c$ is placed to denote terms that disappear in the local Cartesian approximation, $\delta_c \to 0$. We note that due to the current-free condition, $-\frac{\partial B_\phi}{\partial r} = B_\phi/r$. We have let the dispersive terms go to zero and denoted $\wb = \omega - m\Omega (r)$, the Doppler-shifted frequency. For nonzero perturbations, it is clear that the determinant of the coefficients of (\ref{eq:mat1})-(\ref{eq:mat4}) must be zero, giving us a fourth order polynomial equation as our dispersion relation, relating $\ki$ and $\wb$,

\begin{equation}
    \wb^4 + C_3 \wb^3 + C_2 \wb^2 + C_1\wb + C_0 = 0
\end{equation}

where

\begin{align}
    C_3 &= 2\frac{k_\phi \bar k_r \Omega}{k^2} - \frac{\kappa^2}{2\Omega} \frac{k_\phi k_r}{k^2}
    \label{eq:c3}
    \end{align}
    \begin{multline}
    C_2 = -\kappa^2\frac{k_z^2}{k^2} - 2\omega_A^2 \\- \frac{1}{2}\omega_c^2 \delta_c^2 \frac{k^2+k_r\bar k_2}{k^2} + \frac{1}{2}\omega_A \omega_c \delta_c \frac{k_\phi \bar k_r}{k^2}
    \end{multline}
    \begin{multline}
    C_1 = \frac{1}{2}\omega_A \omega_c \frac{\partial \Omega}{\partial \ln r} \delta_c \frac{k^2+k_r\bar k_r}{k^2} + \frac{\kappa^2}{2\Omega}\left(\omega_A^2\frac{k_\phi k_r}{k^2}\right)\\+ \left(\frac{1}{2}\omega_A\omega_c\delta_c\frac{k_z^2}{k^2}\frac{\kappa^2}{2\Omega}\right) - 2\Omega\omega_A^2\frac{k_\phi \bar k_r}{k_z^2} - 2\Omega\omega_A\omega_c \delta_c\frac{k^2+k_r\bar k_r}{k^2}
    \end{multline}
    \begin{multline}
    C_0 = \omega_A^4 + \frac{\partial \Omega^2}{\partial \ln r}\omega_A^2 \frac{k^2+k_r\bar k_r}{k^2} - \frac{1}{2}\omega_A^3 \omega_c \delta_c \frac{k_\phi \bar k_r}{k^2}
    \label{eq:c0}
\end{multline}

and \begin{multline}
\omega_A \equiv \frac{\ki\cdot\B_0}{\sqrt{\mu_0 \rho}} = \frac{1}{\sqrt{\mu_0\rho}}\left( k_z B_z + \frac{m}{r}B_\phi \right), \quad\omega_c \equiv \frac{2B_\phi}{r\sqrt{\mu_0\rho}} \\
k^2 \equiv k_z^2+k_\phi^2-k_r\bar k_r.
\end{multline}

In the limit of $k_{\phi}$, $\delta_c$, $\bar k_r$ $\to$ 0, the local dispersion relation for axisymmetric MRI ~\cite{Balbus1991} is obtained. As we are interested in the largest scale solutions, we will make no assumptions regarding the magnitudes of $k$ or $B$. Figure \ref{fig:disp_rel_2d} shows the WKB solutions for growth rates in the $k$-$V_A$ plane for both pure toroidal (Fig \ref{fig:disp_rel_2d}. a, b) and vertical (Fig \ref{fig:disp_rel_2d}. c, d) magnetic fields. The pronounced effect of the curvature terms is observed best in the toroidal field case, where there are additional differences that arise from the curvature of the magnetic field, but is also subtly observable in the vertical field case. It is important to note that for real disks, both $m$ and $k_z$ are quantized by boundary conditions. For the sake of matching with simulations we will later present, we assume the boundary conditions are periodic both toroidally and vertically, and the disk has height equal to its width $r_2-r_1$ where $r_2$ and $r_1$ are the outer and inner radii. This gives us a minimum value for $k_z$ of $k_1 \equiv 2\pi /2(r_2-r_1)$. Solutions matching the vertical periodic boundary condition are integer multiples $k_{z_n} = nk_1$. 

Due to the variance of equilibrium quantities in the radial direction, $k_r$ is a very poor choice for a quantum number, as radially sinusoidal eigenfunctions do not portray the fact that flow shear and field-line bending change as a function of radius. Additionally, one would expect waves propagating in a disk to move with the disk, but such eigenfunctions would be broken up by flow shear. This motivates the construction of a better model for this system which utilizes an ordinary differential equation to find the eigenfunctions and complex frequencies through a shooting method.

\section{Global stability analysis and the energy principle } \label{sec:global}

Now we examine the linear global stability of our system by allowing the radial variation of the linear perturbations. In the ideal limit (in the absence of magnetic and fluid diffusivities), the velocity and magnetic perturbations are expressed in terms of the displacement vector $\x(r, \phi, z, t) = [\xi_r(r),\xi_{\phi}(r),\xi_z(r)]\exp{i(m\phi+kz - \omega t)}$  (\cite{Chandrasekhar_2006}), as $ {\bf \vt} = -i\wb\x - r\frac{\partial \Omega}{\partial r} \xi_r \e_\phi$ and ${\bf \Bt} = i\left(\ki\cdot\B\right)\x+\frac{2 B_\phi}{r}\xi_r \e_\phi$, respectively. 

Combining with incompressibility,
   $ \nabla\cdot\x = \frac{1}{r}\frac{\partial}{\partial r}\left( r\xi_r\right) + \frac{im}{r}\xi_\phi + i k \xi_z = 0$, we obtain the components of the linearized momentum equation,
   \begin{multline}
 \left(-\wb^2 + \omega_A^2 + \omega_s^2 + \omega_c^2 \delta_c\right) \xi_r +\left[i\omega_A\omega_c \delta_c+ 2i\wb \Omega(r)\right] \xi_{\phi} \\= - \frac{\partial \widetilde P}{\partial r}\label{eq:mri1}
 \end{multline}
 \begin{multline}
\left(-\wb^2 + \omega_A^2\right) \xi_{\phi}  -i \left[2\wb \Omega(r) + \omega_A \omega_c \delta_c\right] \xi_{r} \\= \frac{im \widetilde P}{r}
\label{eq:mri2}
\end{multline}
\begin{equation}
\left(-\wb^2 + \omega_A^2\right) \xi_{z} = -i k_z \widetilde P,
\label{eq:mri3}
\end{equation}

 where $\widetilde P \equiv \frac{1}{\rho}\left(\widetilde{p} + \frac{1}{\mu_0}{\bf \Bt}\cdot \B_{0}\right) $, and $\omega_s^2 \equiv \frac{\partial \Omega^2}{\partial \ln r}$.
By combining (\ref{eq:mri1}-\ref{eq:mri3}), an ordinary differential equation is then obtained, 

\begin{multline}
\label{eq:ode}
    \left[ \frac{\left(\wb^2 - \omega_A^2\right)}{k^2x+m^2}xu'\right]' + u \left[  m \left(\frac{\Omega\wb + \frac{1}{2}\omega_A\omega_c}{k^2x+m^2}\right)'\right. \\ \left. - \frac{(\wb^2 - \omega_A^2 - \omega_s^2 - \omega_c^2)}{4x} + \frac{\left(\Omega\wb + \frac{1}{2}\omega_A\omega_c\right)^2}{k^2x+m^2}
     \frac{k^2}{\wb^2-\omega_A^2}\right] = 0
\end{multline}

where we have defined
$u \equiv r\xi_r $ and $x \equiv r^2$.
In the limit of $\omega_c \rightarrow 0$, Eq.~\ref{eq:ode} is reduced to the ordinary differential equation found in \cite{Khalzov_2006} with only vertical magnetic field. The variable substitutions here were made for compactness and simplicity, though the equation in terms of the original variables is easily recovered for comparison with other work. It should be mentioned that other forms of differential equations for rotating plasmas \citep{hameiri1981}, local pressure-driven instabilities ~\citep{bondeson1987}, current carrying disks~\citep{Ebrahimi_2011_mom_tran}, and for more general astrophysical disk equilibrium~\citep{keppens2002,blokland2005magneto} were also previously obtained.  Here, we solve this equation using a complex shooting method, employing a zero-finding algorithm to obtain the eigenfunctions and their eigenvalues $\bar \omega$ (the sign of whose imaginary part signifies instability) as a function of $k$, flow shear, $m$, and $\B_0$. 
In this paper we will focus mostly on the lowest non-axisymmetric values of $k$ and $m$, though this method could enable study on higher modes, different configurations of mixed fields, and different flow profiles in the future. Before we present the numerical global solutions of Eq.~\ref{eq:ode}, we first revisit the energy principle for a general global stability criterion.

The stability of static ideal plasmas can generally be determined using the ideal MHD energy principle {\color{red}{\citep{bernstein1958,freidberg_rwm,schnack2009}}} without solving differential equations. The stability criterion is constructed based on the sign of energy integral, the volume integral $\int {\x \cdot F_0(\x) dx^3}$, where $F_0(\x)$ is the self-adjoint static force operator obtained from the linearized momentum equation.
However, in the presence of an equilibrium mean flow, self-adjointness of the linear stability problem is lost, and the stability criterion was reconstructed with the non-zero flows by \cite{frieman_rotenberg}. In flowing plasma, the total force operator in terms of Lagrangian displacement (i.e. the displacement of a fluid element moving with the equilibrium flow) was given as $F (\x)= F_0(\x) + F_1(\x) = F_0(\x) + \nabla \cdot [\rho \x (\bar{\vi} \cdot \nabla) \bar{\vi} -\rho \bar{\vi}(\bar{\vi}\cdot \nabla) \x]$, where the modification due to flows has been introduced through $F_1(\x)$ with the two additional flow-dependent terms. The sufficient stability condition (for  $\x$ not to have a positive growth) was found to be $\int {\x^{\ast} \cdot F(\x) dx^3}<0$~\citep{frieman_rotenberg}. For our differentially rotating system here, we expand $F_1(\x)$, 
\begin{equation}
    F_1(\x) = \nabla \cdot \bf\Gamma =\nabla \cdot [\rho \x (\bar{\vi} \cdot \nabla) \bar{\vi} -\rho \bar{\vi}(\bar{\vi}\cdot \nabla) \x]
\end{equation}
where in curvilinear cylindrical coordinate with only an azimuthal flow, it reduces to 
\begin{multline}
        F_1(\x) = \nabla \cdot \bf{\Gamma} = [\frac{1}{r} \frac{\partial}{\partial r} (r \Gamma_{rr}) -\frac{\Gamma_{\phi\phi}}{r}]\hat{r} + [\frac{1}{r} \frac{\partial}{\partial \phi} ( \Gamma_{\phi \phi})]\hat\phi\\
    =\rho \left[-\frac{d\Omega^2}{d lnr}\xi_r - \frac{v_{\phi}^2}{r^2} (r\xi_r)^{\prime} -i k_z \frac{v_{\phi}^2}{r} \xi_z\right]\hat{r} - \rho \frac{v_{\phi}^2}{r^2}  (im) \xi_r \hat{\phi}
    \label{eq:force1}
    \end{multline}

Combining Eq.~\ref{eq:force1} with incompressibility,
   $ \nabla\cdot\x = \frac{1}{r}\frac{\partial}{\partial r}\left( r\xi_r\right) + \frac{im}{r}\xi_\phi + i k \xi_z = 0$, the total energy integral of \cite{frieman_rotenberg}  due to flow-dependent terms becomes;
   
   \begin{multline}
      \int  \x^\ast \cdot F_1(\x) dx^3 = \delta w_I +\delta w_{II} = \\ \int \rho(-  \frac{\partial\Omega^2}{\partial lnr}|\xi_r|^2  + i m \Omega^2(r) [\xi_{\phi}\xi_r^\ast - \xi_{\phi}^\ast \xi_r])dx^3
      \label{eq:dw}
   \end{multline}
   Interestingly, the first term on the RHS of the Eq.~\ref{eq:dw} is the differential rotation term, which similarly to the local WKB dispersion relation ($C_0<0 $ in Eq.~\ref{eq:c0}) results in the MRI instability. The first term causes instability in systems with angular velocity a globally decreasing function of radius. We note that the additional second term on the RHS however, only arises due to the effect of spatial curvature (here in cylindrical geometry), and can be obtained in terms of the mean equilibrium flows. By combining Eq.~\ref{eq:mri2} , Eq.~\ref{eq:mri3} and the incompressibility, we obtain 
   \begin{multline}
      \xi_{\phi} = \frac{r^2 k_z^2}{k_0^2}[\frac{2i\Omega(r) \bar{\omega}}{(\omega_A^2 -\bar{\omega}^2)}\xi_r + \frac{i\omega_A \omega_c\delta_c}{(\omega_A^2 -\bar{\omega}^2)}\xi_r \\ + \frac{im}{r^2k_z^2}(r\xi_r)^{\prime}]
      \label{eq:xip}
       \end{multline}
       where, $k_0^2= (r^2k_z^2+m^2)$, inserting this equation in Eq.~\ref{eq:dw}, we find 
       \begin{multline}
           \delta w_{II} = \int [m\Omega^2 (r) \frac{ - 2\rho r^2k_z^2}{k_0^2}\frac{(2m\Omega(r)\bar{\omega}  +\omega_A\omega_c\delta_c)}{(\omega_A^2 -\bar{\omega}^2)}|\xi_r|^2 \\ -\rho\frac{m^2\Omega^2}{k_0^2}\left((r\xi_r)^{\prime}\xi_r^\ast + \xi_r (r\xi_r^\ast)^{\prime})\right]dx^3
           \label{eq:dw2}
       \end{multline}
       
  This introduces yet another source of free energy for the instability due to nonzero flows,  i.e. causes a positive contribution to  $\int {\x^{\ast} \cdot F(\x) dx^3}$ when 
  $\omega^2_A  < \bar{\omega}^2$. This term (Eq.~\ref{eq:dw2})  is destabilizing  due to 1) nonzero non-axisymmetric modes (i.e. $m\neq 0$) 2) nonzero averaged global angular velocity and magnetic curvature. 
   
   Alternatively, in addition to considering the Frieman \& Rotenberg energy term (eq.~\ref{eq:dw}), we can directly calculate the energy terms from the sets of momentum equations (eq.~\ref{eq:mri1}-\ref{eq:mri3}). Multiplying Eqs. eq.~\ref{eq:mri1}-\ref{eq:mri3} by $\xi_r^\ast$,  $\xi_\phi^\ast$ and $\xi_z^\ast$, respectively and integrating over the volume we obtain, 
   
   \begin{multline}
        \int  [ (\omega_A^2 -\bar{\omega}^2) |\xi|^2  + \omega_s^2 |\xi_r|^2 +\\ 2 i\bar{\omega} \Omega(r) (\xi_{\phi}\xi_r^\ast - \xi_{\phi}^\ast \xi_r)] r dr =0 
   \end{multline}
   
   In the absence of flow, using the incompressibility condition and eq.~\ref{eq:xip}, from this equation the general $\delta w$ condition for a current-free cylinder  with magnetic field is obtained $\int \omega_A^2 [|(r \xi_r)^{\prime}|^2/k_0^2 +|\xi_r|^2] r dr $ (\cite{schnack2009}). With flow and non-axisymmetric perturbations, additional terms from Eq. 19 contribute to the energy. The contributions of the individual  terms in the energy equation will be examined in a future study.

\subsection{Global solutions with vertical magnetic field}\label{sec:global_z}
\begin{figure}
    \centering
        (a)
        \includegraphics[width=\columnwidth]{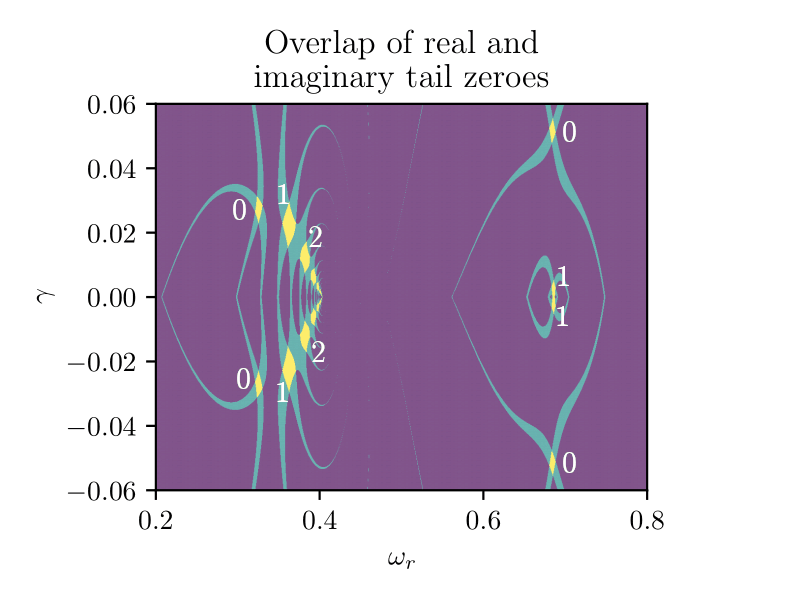}
        (b)
        \includegraphics[width=\columnwidth]{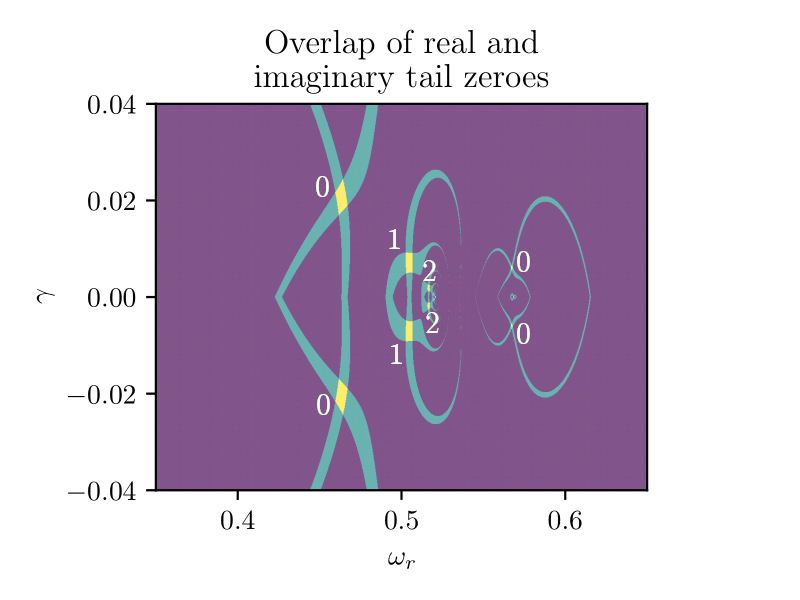}
    \caption{Locations of the real and imaginary shooting tails and their overlap, where a solution fitting the boundary conditions exists. Purple indicates neither the real nor the imaginary tail is within a tolerance of zero; teal indicates one tail is near zero; yellow indicates both tails go to zero and there is a mode in that neighborhood. (a) Shows the modes at $k=2 k_1$, $V_A = 0.2$ $r_1 \Omega_0$, and (b) shows the modes at $k=k_1$, $V_A=0.57$ $r_1 \Omega_0$. The left-hand family of solutions is the low-frequency (outer) curvature mode, and the right-hand family of solutions is the high-frequency (inner) MRI mode. Both figures use solely vertical background field and $m=1$. Numbers show $n_r$.}
    \label{fig:modes}
\end{figure}

\begin{figure}
    \centering
        (a)
        \includegraphics[width=\columnwidth]{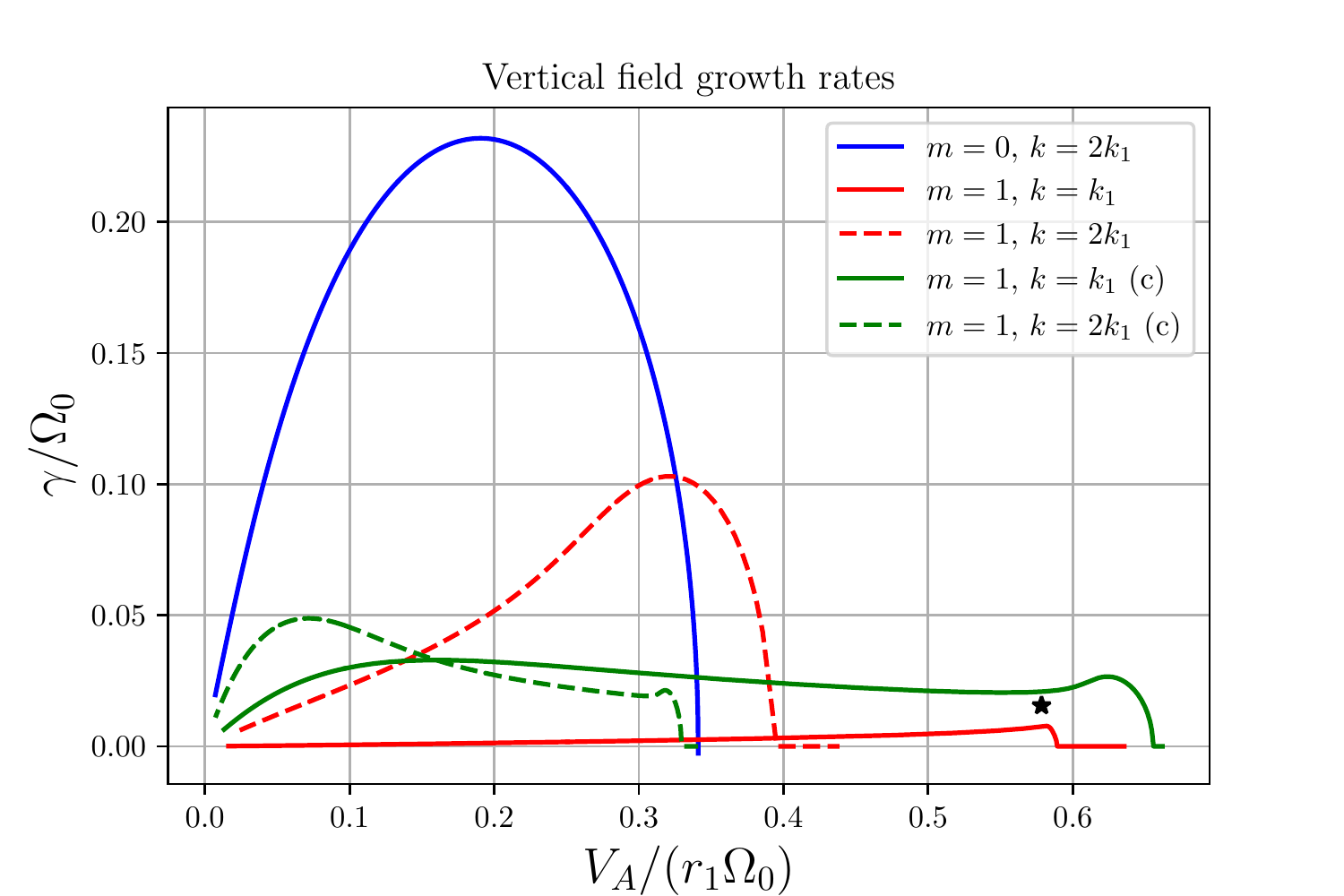}
        (b)
        \includegraphics[width=\columnwidth]{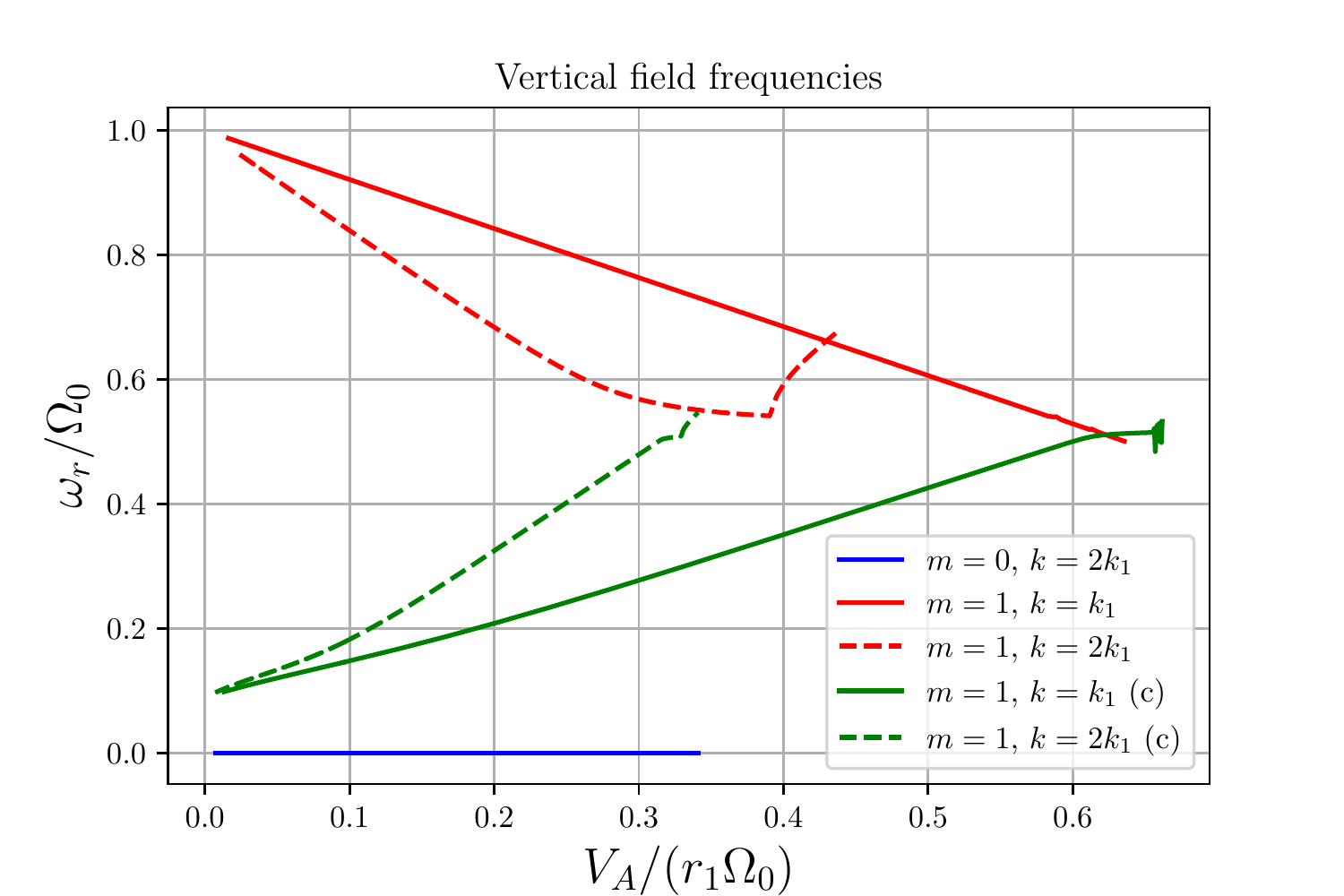}
    \caption{(a) Growth rates and (b) mode frequencies from global shooting method with vertical seed field,  for axisymmetric ($m=0$) MRI, and non-axisymmetric ($m=1$) MRI and curvature modes with, $k=k_1, 2k_1$. Nonlinear NIMROD simulation (NVF-C in table 1) is performed at the star point.}
    \label{fig:Bz_growthrates}
\end{figure}

\begin{figure}
    (a) MRI \hspace{35mm} (b) curvature \\
    \includegraphics[width=0.49\columnwidth]{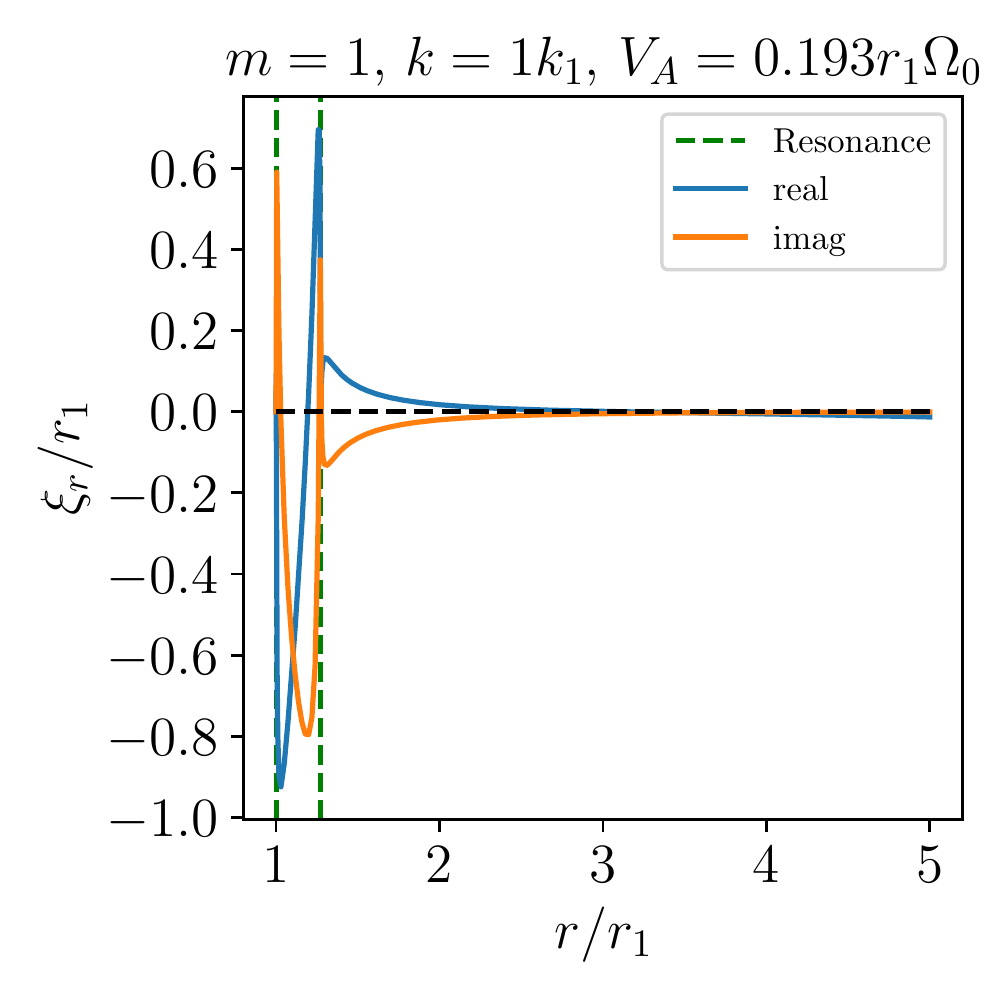}
    \includegraphics[width=0.49\columnwidth]{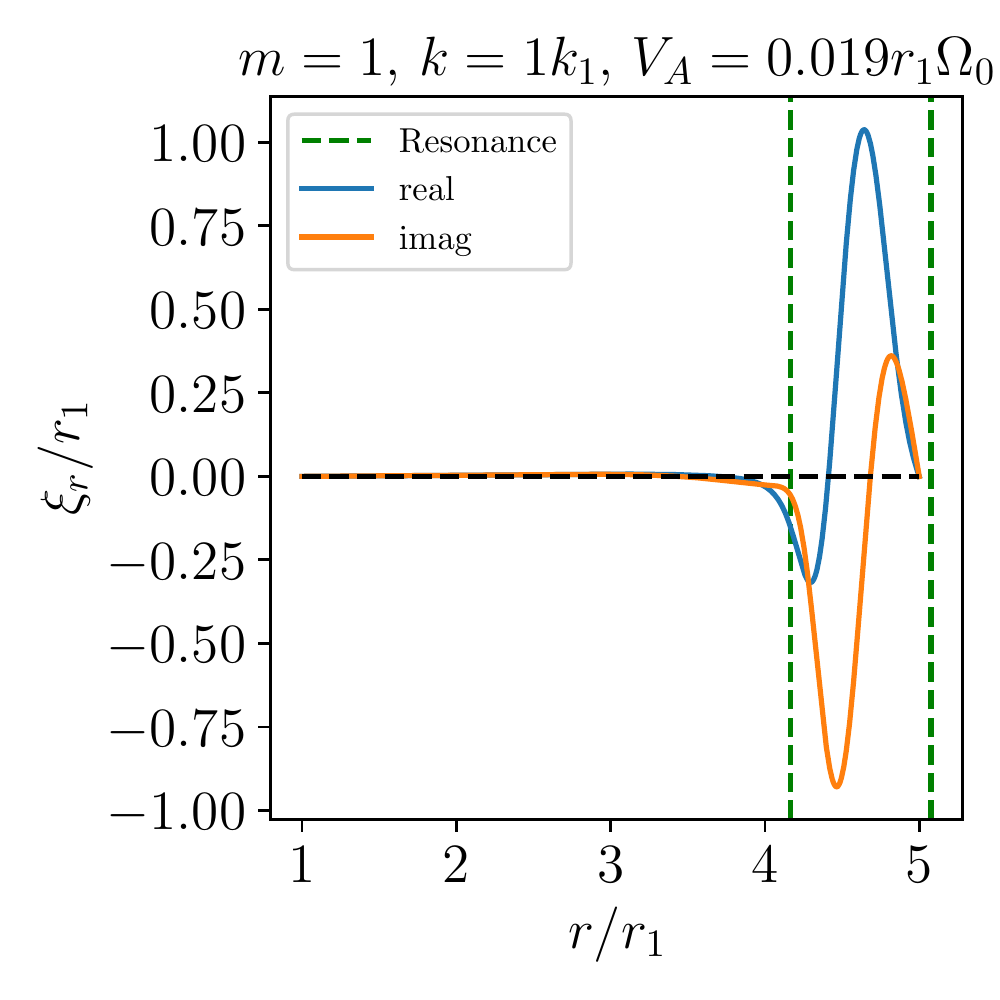}\\
    (c) MRI \hspace{35mm} (d) MRI \\
    \includegraphics[width=0.49\columnwidth]{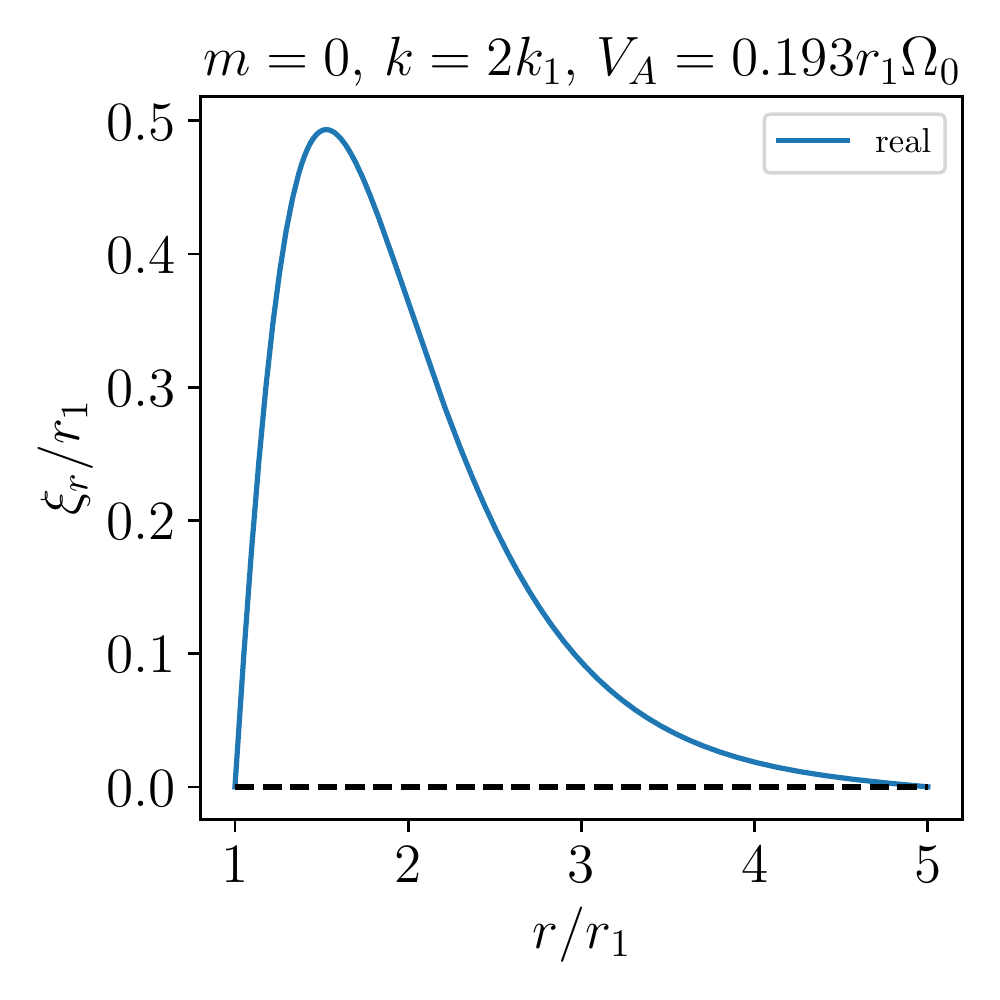}
    \includegraphics[width=0.49\columnwidth]{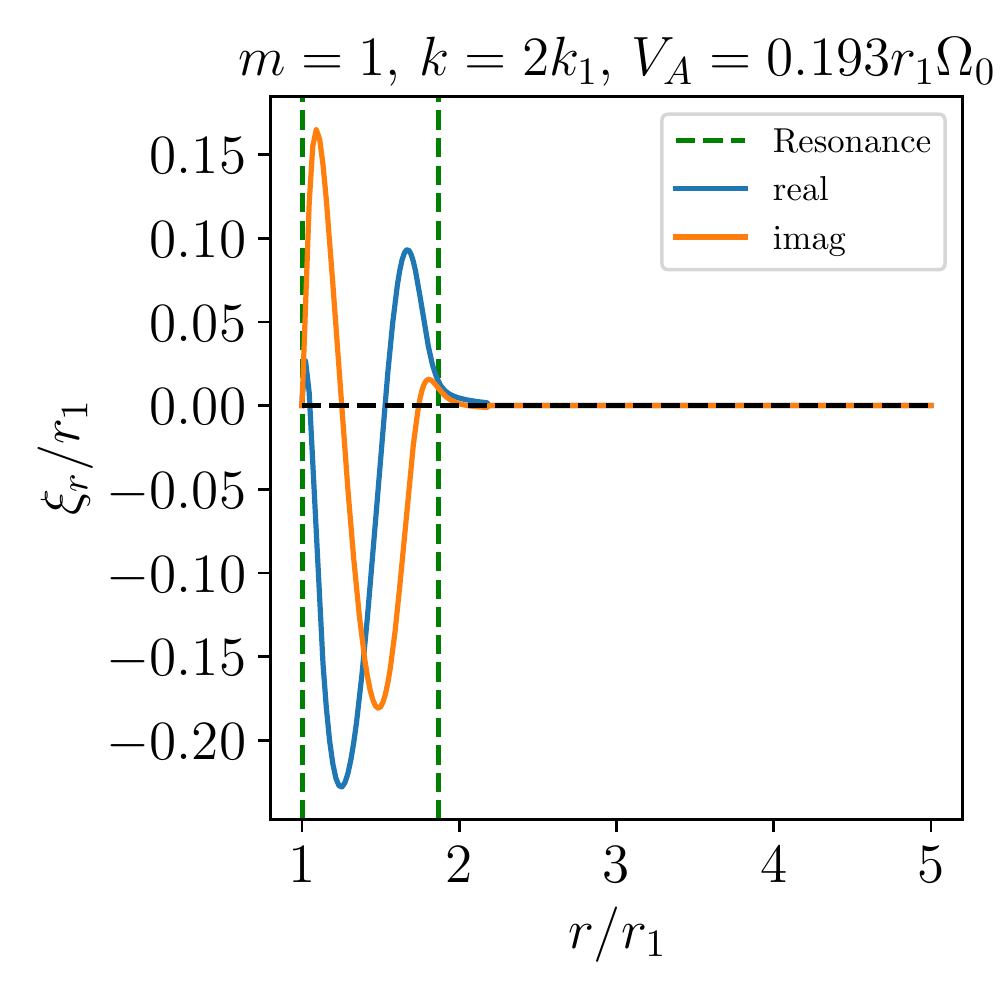}\\
    (e) curvature \hspace{28  mm} (f) most global curvature\\
    \includegraphics[width=0.49\columnwidth]{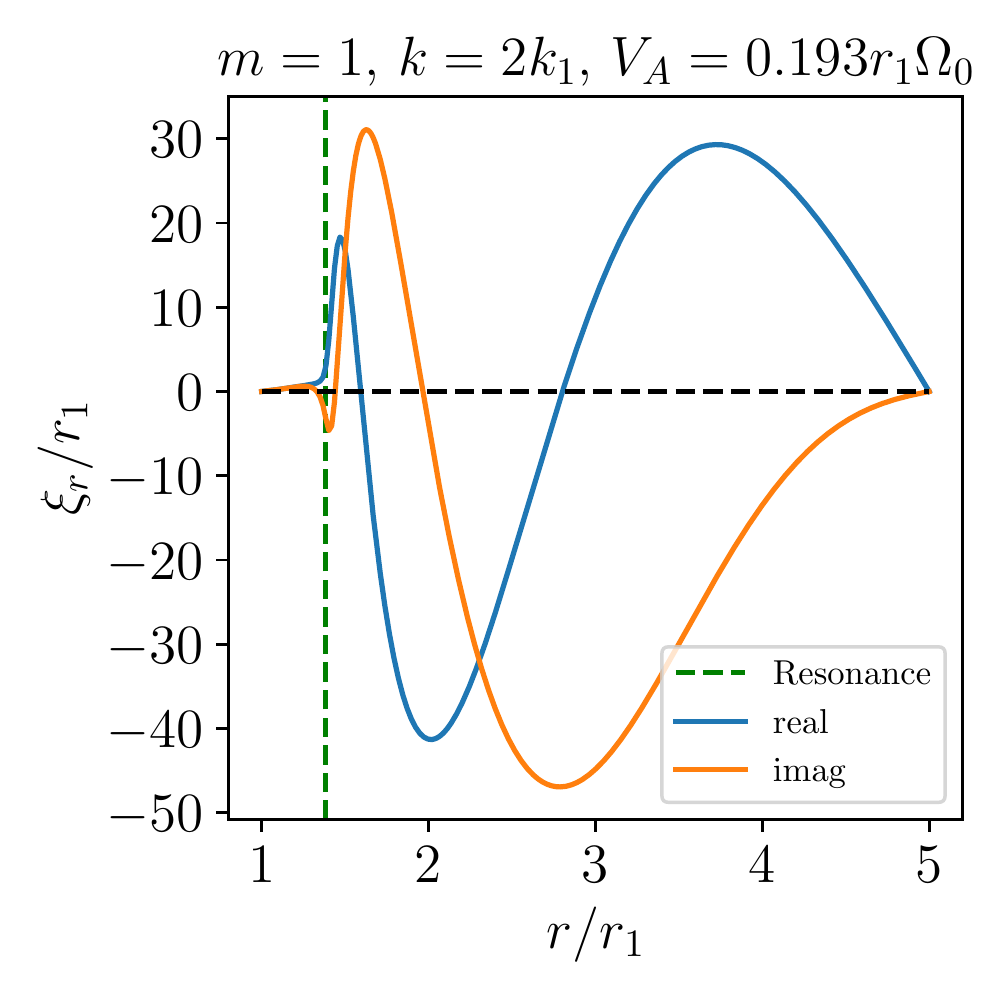}
    \includegraphics[width=0.49\columnwidth]{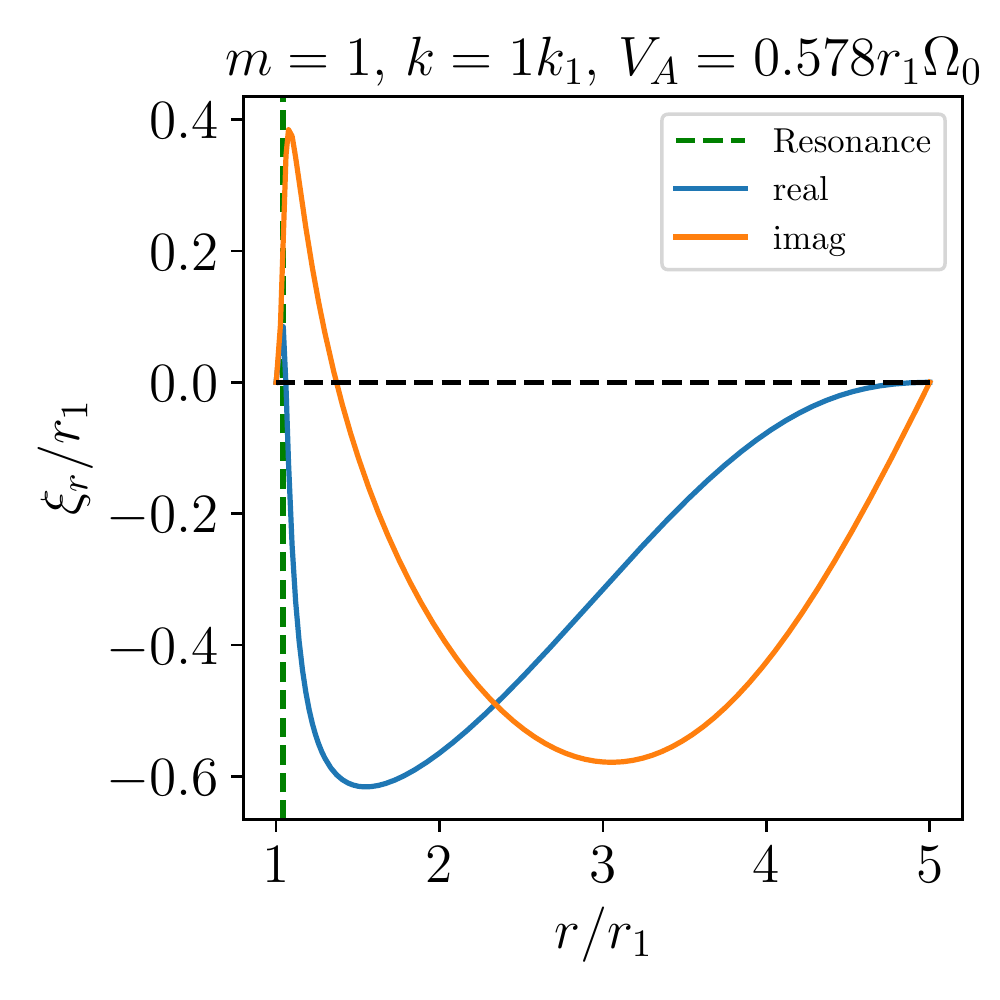}
    \caption{Mode structures for vertical magnetic field from shooting method. (a) and (b) m=1 mode at low field showing MRI and curvature mode localization between Alfvénic points at the inside/outside of the disk respectively; (c) and (d), (e) at intermediate field in Fig. \ref{fig:bifur} at (c) point 6, the point of convergence, (d) point 7, the $m=1$ MRI mode, and (e) point 1, the $m=1$ curvature mode (outer Alfv\'enic resonance not pictured at $r_{out}=20.3r_1$). (f) at high field for the most global curvature mode for m=1 $k=k_1$ ($\gamma=0.021,\omega_r=0.485$ and Alfv\'enic resonances at $r_{in}=1.04r_1$, $r_{out}=10.3r_1$) }
    \label{fig:eigenfuncs}
\end{figure}

First, we present the solutions obtained from Eq.~\ref{eq:ode} with only the vertical magnetic field and Keplerian flow. We use a complex shooting method in which we consider the solution at the boundary (to which we refer as the complex and imaginary 'tails') as a function of complex frequency $\omega$ (see Fig.~\ref{fig:modes}). The modes are obtained when $\xi$ goes to zero at the right-hand boundary, enforcing the boundary condition. The growth rates and frequencies of several unstable m=0 and m=1 modes with positive $\gamma$ vs the normalized Alfvén velocity ($V_A/r_1 \Omega_0$) are shown in Fig.~\ref{fig:Bz_growthrates}. As expected with only the vertical magnetic field, the most unstable MRI modes are the axisymmetric modes, which span from weak to strong fields. However, we find two sets of distinct non-axisymmetric $m=1$ instabilities with different mode structures. The first, which resemble usual localized MRI modes (inner modes), centered about the point of maximum flow shear, we refer to as such; the second, which can feature more global structure at lower magnetic fields, we refer to as curvature modes (outer modes). The MRI and curvature modes for $m=1$, $k=k_1$, at small magnetic fields are shown in Fig.~\ref{fig:eigenfuncs} (a) and (b), respectively.

As seen in Fig.~\ref{fig:Bz_growthrates}, we also uncover that the new \emph{global} curvature mode remains unstable to much higher magnetic fields. The non-axisymmetric modes, as discussed in \cite{tajima_shibata_1997} and \cite{Ogilvie_1996}, are localized between the two Alfvén resonance points where 

\begin{multline}
    \Re(\bar\omega)^2-\omega_A^2 = \left(\omega_r-m\Omega\right)^2 - \omega_A^2 =\\ \left(\omega_r-m\Omega - \omega_A\right)\left(\omega_r-m\Omega + \omega_A\right)=0. 
    \label{eq:resonance_cond}
\end{multline}

These are the points where the magnitude of the Doppler-shifted frequency of the mode is equal to its Alfvén frequency, and are always centered about the point of corotation, where $\omega_r = \Omega(r)$. In particular, localized MRI and curvature modes in systems with weak magnetic fields (i.e. super-Alfvénic flows) feature structures localized in the inner and outer parts of the system as also obtained by \cite{Ogilvie_1996} { and \cite{goedbloed2022}}, due to the differing locations of the Alfvén singularities. Inner (MRI) modes are located between the conducting boundary at $r_1$ and the singularity $r_{out}$; outer (curvature) modes are between the singularity $r_{in}$ and the conducting boundary at $r_2$ (Figs.~\ref{fig:eigenfuncs} (a) and (b)). We find that increasing the magnetic field results in both modes taking on more global properties, though this is especially and more rapidly true for the curvature modes. Global versions of curvature mode can be seen in Figure \ref{fig:eigenfuncs} (e) and (f) with increasing normalized Alfvén flows. The comparison of mode structures for MRI and and the curvature mode with the 
exactly the same sets of parameters ($k=2k_1$ and $V_A=0.193 r_1\Omega_0$) can also be seen in Figs.~\ref{fig:eigenfuncs} (d) and (e). The frequencies of these two modes also exhibit differing properties. For low fields, each mode's frequency reflects the angular velocity of the system at the mode's local position, and converge on a moderate frequency as fields grow. This can be seen in Figure \ref{fig:Bz_growthrates} (b), where the MRI mode's frequency in the low-B limit converges to $\Omega_0$, the system's angular velocity at the inner wall, and the curvature mode's frequency similarly converges to $\Omega_0(r_1/r_2)^{3/2}$ $(0.0894\Omega_0\text{ in our system})$, the angular velocity at the outer wall.


\begin{figure}
    (a) high field \hspace{30mm}  (b) low field\\
    \includegraphics[width=0.45\columnwidth]{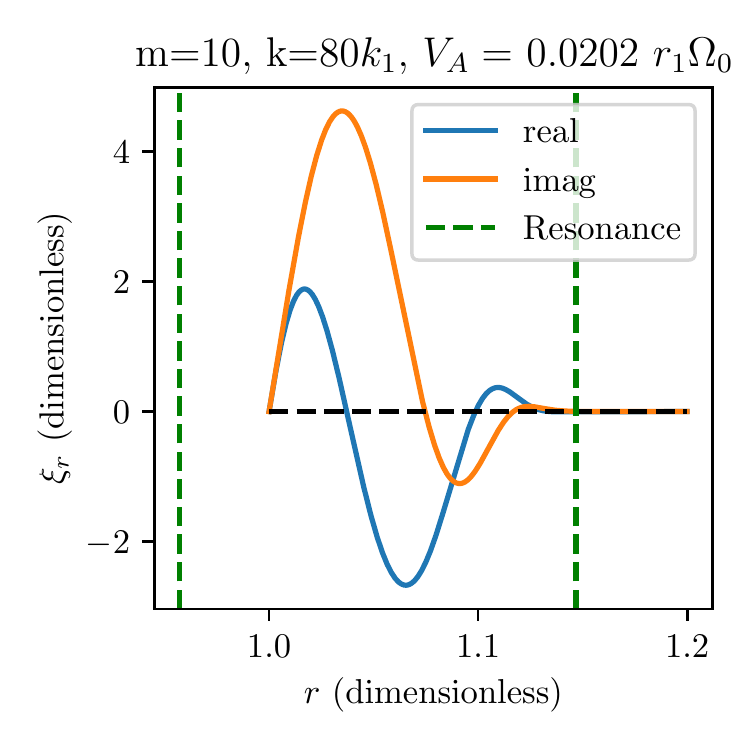}
    \includegraphics[width=0.45\columnwidth]{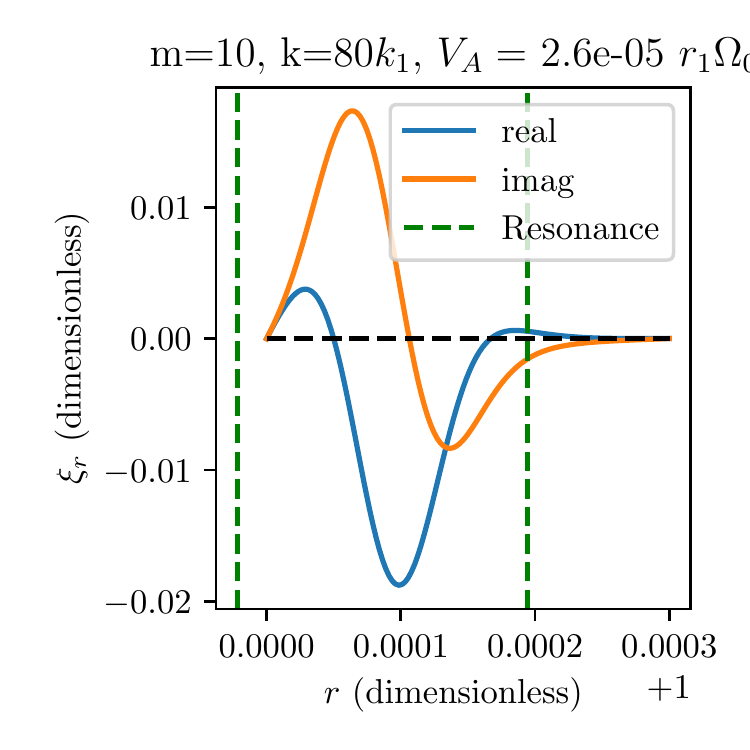}\\
    \hspace{35mm}{ (c)}\\
    \includegraphics[width=0.9\columnwidth]{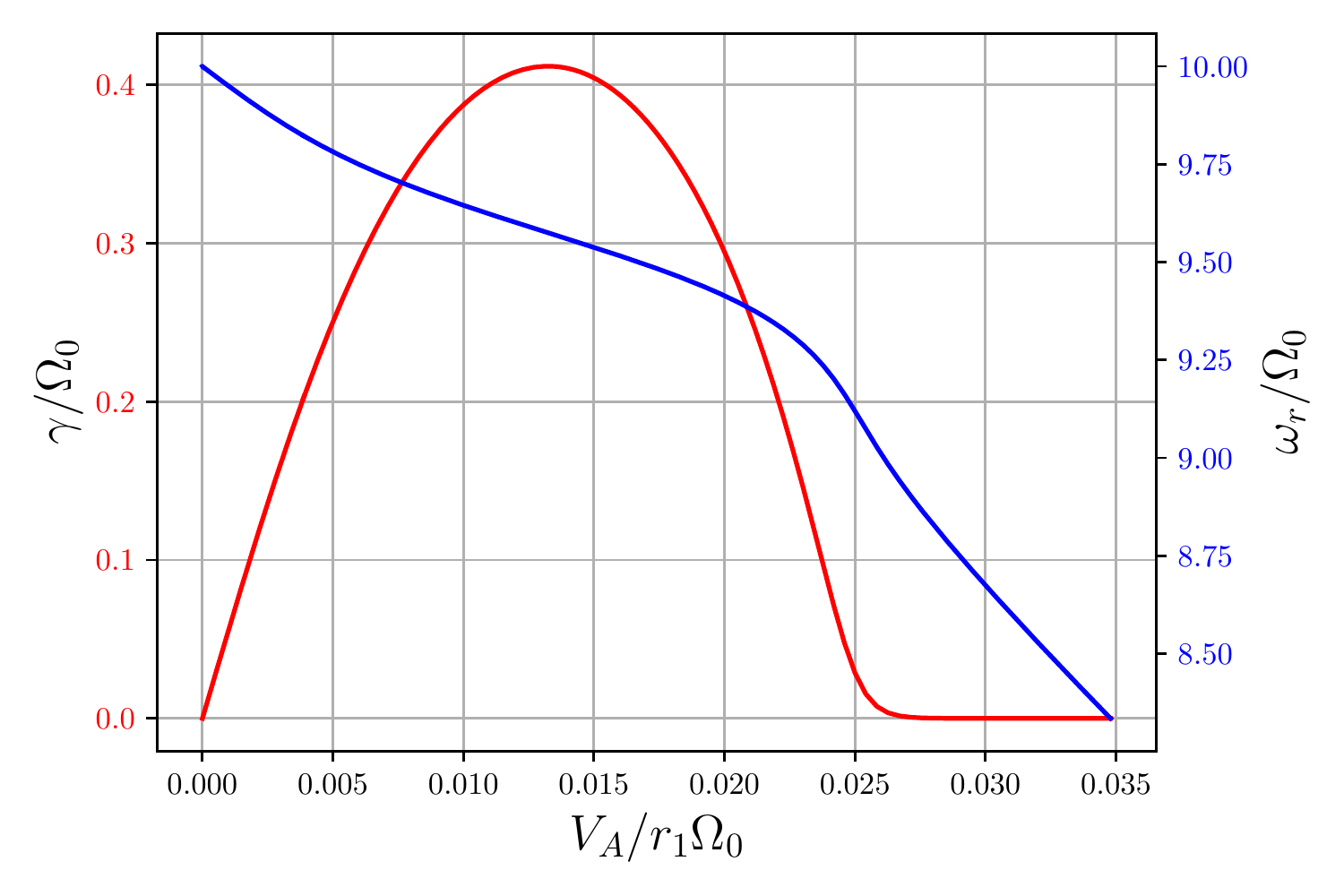}
     \caption{(a) high field and (b) low field modes shown here in the super-Alfv\'enic region, i.e. high-$m$ ($m=10$). We find that at high toroidal mode number and high $k$ ($k=80k_1$), the range of magnetic fields to which the system is linearly unstable to a perturbation is narrowed to the low field region. This can be seen in comparison with Figs.~\ref{fig:Bz_growthrates}, \ref{fig:gaplimit}, and \ref{fig:bp_growthrates} varying $V_A$ for $m=1$.}
    \label{fig:goedbloed_eigenfuncs}
\end{figure}

We also search in the limit of super Alfv\'enic flows where $\omega_A \ll
m\Omega$ as studied by \cite{Ogilvie_1996} and \cite{goedbloed2022}. In this region, the high-$m$, high-$k$ non-axisymmetric Alfv\'enic resonance instability is extremely localized due to being confined between two very close-by Alfv\'enic resonances. This can be seen in Fig.~\ref{fig:goedbloed_eigenfuncs} with $m=10$ and $k=80k_1$. The complex eigenvalues are similar to their approximate counterparts in \cite{goedbloed2022}. Regardless of the use of a global treatment, these extremely localized instabilities only occupy a narrow region of parameter space, as they are only unstable for very weak magnetic fields (i.e. super-Alfv\'enic flows). For the remainder of this paper, we will focus on the most global non-axisymmetric instabilities that are unstable in a far wider range of magnetic fields, from small fields to fields with corresponding Alfv\'en velocities on the order of $r_1\Omega_0$.

To further elucidate the distinct nature of MRI and curvature modes, we also present the complex solutions in a 2D plane in Fig.~\ref{fig:modes}, where they can be seen as two discrete families of solutions: a low-frequency (the structures seen to the left) set of curvature modes and a high-frequency set of MRI modes (the structures seen to the right), similar to plots seen in \cite{Khalzov_2006} and \cite{goedbloed2022}.We show the solutions in the low- and high-field limits, at two different values of $k$ in Fig.~\ref{fig:modes} (a),(b), respectively. We primarily illustrate the properties of the most unstable solutions.
For $m=0$, these modes are characterized by the number of times they cross the $x$-axis ($n_r$); however, non-axisymmetric modes are necessarily complex, and therefore this definition loses its meaning, though we may keep this labeling convention defined by the value of $n_r$ each mode takes on when quasi-continuously changed from $m=1$ to $m=0$. We may additionally verify the applicability of equation \ref{eq:resonance_cond}; for the low field case in Fig.~\ref{fig:modes} (a), the Alfvénic points are calculated based on the inner and outer angular velocities. In the MRI (inner) case, the two Alfvénic points are located at $\omega_{r\pm(inner)} = 0.68,$ $1.31.$ In the curvature (outer) case, the two Alfvénic points are $\omega_{r\pm(outer)} = -0.22,$ $0.40.$ This gives two distinct ranges for the frequencies of inner and outer modes, which can be seen in the two fractal structures in Fig.~\ref{fig:modes} (a). This also holds for the higher field case in Fig.~\ref{fig:modes} (b), though care must be taken as all modes in the high field case are more global than their lower field counterparts and therefore $\Omega$ is an intermediate as opposed to a boundary value. Therefore, the range of real frequencies in these modes is consistent with equation \ref{eq:resonance_cond}, as modes with different frequencies are located at different radial points with different values of $\Omega$. 

\begin{figure}
    \centering
        \includegraphics[width=\columnwidth]{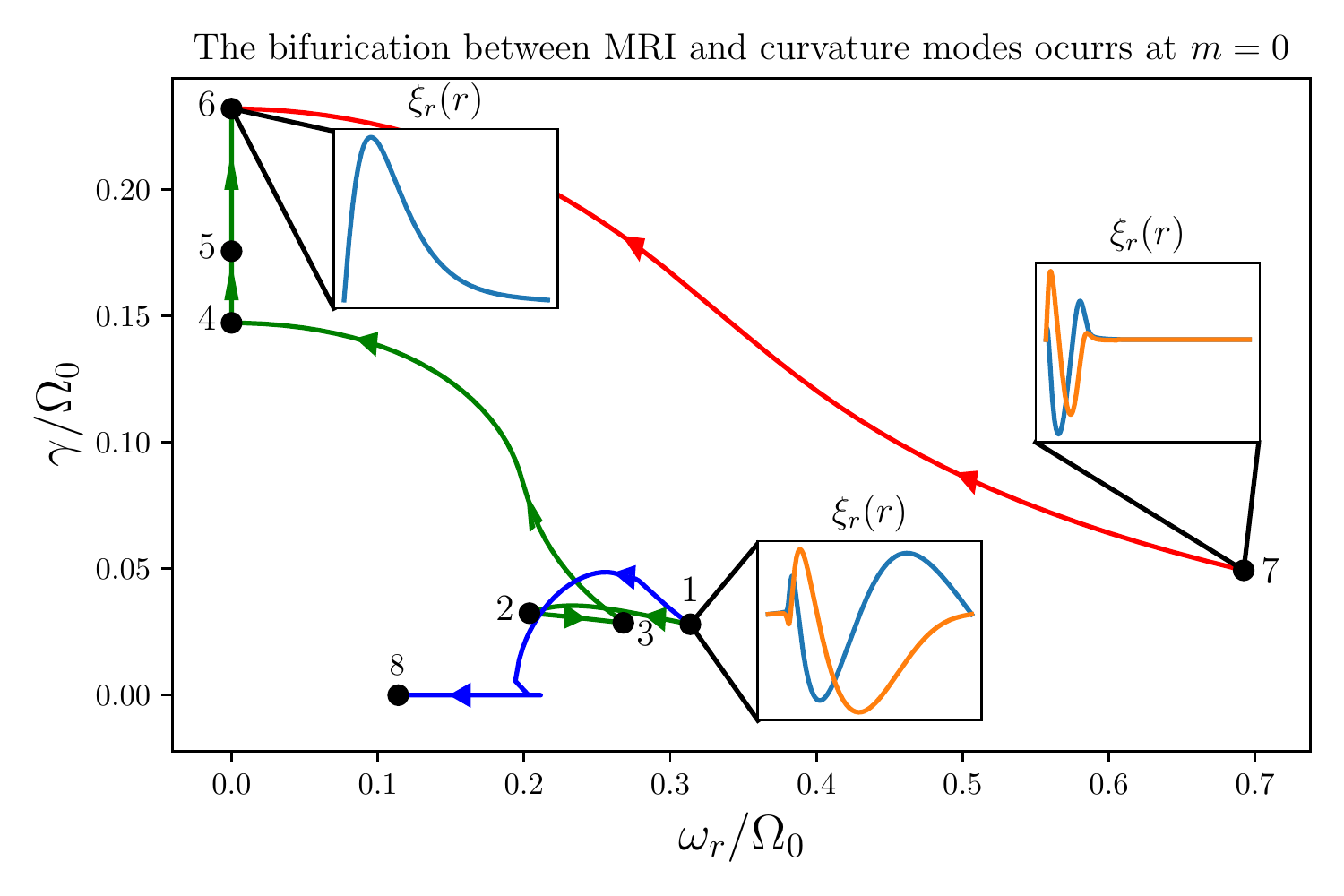}
        \centering
        (a)

    \vspace{5mm}
    
    (b) Points labeled in (a).
        \begin{tabular}{|c|c|c|c|}
        \hline
        Point & $m$ & $k/k_1$ & $V_A/(r_1\Omega_0)$\\
        \hline
        1 & 1 & 2 & 0.193 \\
        2 & 1 & 1 & 0.193 \\
        3 & 1 & 1 & 0.289 \\
        4 & 0 & 1 & 0.289 \\
        5 & 0 & 2 & 0.289 \\
        6 & 0 & 2 & 0.193 \\
        7 & 1 & 2 & 0.193 \\
        8 & 0 & 2 & 0.193 \\
        \hline
        \end{tabular}
    \caption{Progression through the   $m$, $k$, $V_A$ continuum between the curvature mode (point 1) and the MRI mode (point 7). Two distinct mode structures are recognized with the same $m$, $k$, $V_A$ at points 1 and 7. Direct traversal from points 1 and 7 to $m=0$ resulted in stability, partially seen in the blue curve ending at point 8, so an indirect path through varying $k$ and $V_A$ was taken in points 2-6.}
    \label{fig:bifur}
\end{figure}

The eigenfunctions and eigenfrequencies exist in a phase space defined by considering $m$, $k$, and $V_A$ as continuous quantities, though the first two are discrete in a real system. To further explore the main differences between the non-axisymmetric MRI and curvature modes, we perform an exercise by gradually varying $m=1\to m=0$ starting from the same state (i.e $m=1$, $k=2k_1$, $V_A = 0.193 r_1\Omega_0$). This is shown in Fig.~\ref{fig:bifur}. We find that both modes will transition to a global axisymmetric state as shown in Fig.~\ref{fig:eigenfuncs} (c). This shows that one mode can be obtained from the other by continuously varying azimuthal wavevector. However, there is only a unique path for the local $m=1$ MRI mode (point 7) to transition to the $m=0$  MRI mode (as shown in red in Fig.~\ref{fig:bifur}), while the transition from the non-axisymmetric curvature mode (point 1) is non-trivial. Immediately varying $m$ from point 1 results in the mode becoming stable in between $m=1$ and $m=0$ (this trajectory is shown in blue). This is avoided by first changing $k$ and $V_A$ as detailed in the table Fig.~\ref{fig:bifur} (b). This path, as shown in green, also results in an axisymmetric MRI mode, which, upon varying $k$ and $V_A$ back to their original values, converges with the mode obtained from point 7. This point of convergence is at point 6. 

To further corroborate the hypothesis that the curvature mode is driven by the global derivative terms in the WKB analysis, we will now examine what happens as we change the size of our system (preserving the cross-sectional dimensions $h/(r_2-r_1)$). As the aspect ratio $r_1/(r_2-r_1)$ of our system grows, the radially dependent terms vary increasingly little across our system and the average curvature of motion goes down with $1/r_1$, leading to a quasi-cartesian approximation. Conversely, upon shrinking the system aspect ratio, radially dependent terms vary increasingly more and the average curvature of the flow grows asymptotically, leading to a system dominated by the curvature terms. 

\begin{figure}
    \centering
        \centering
        \includegraphics[width=\columnwidth]{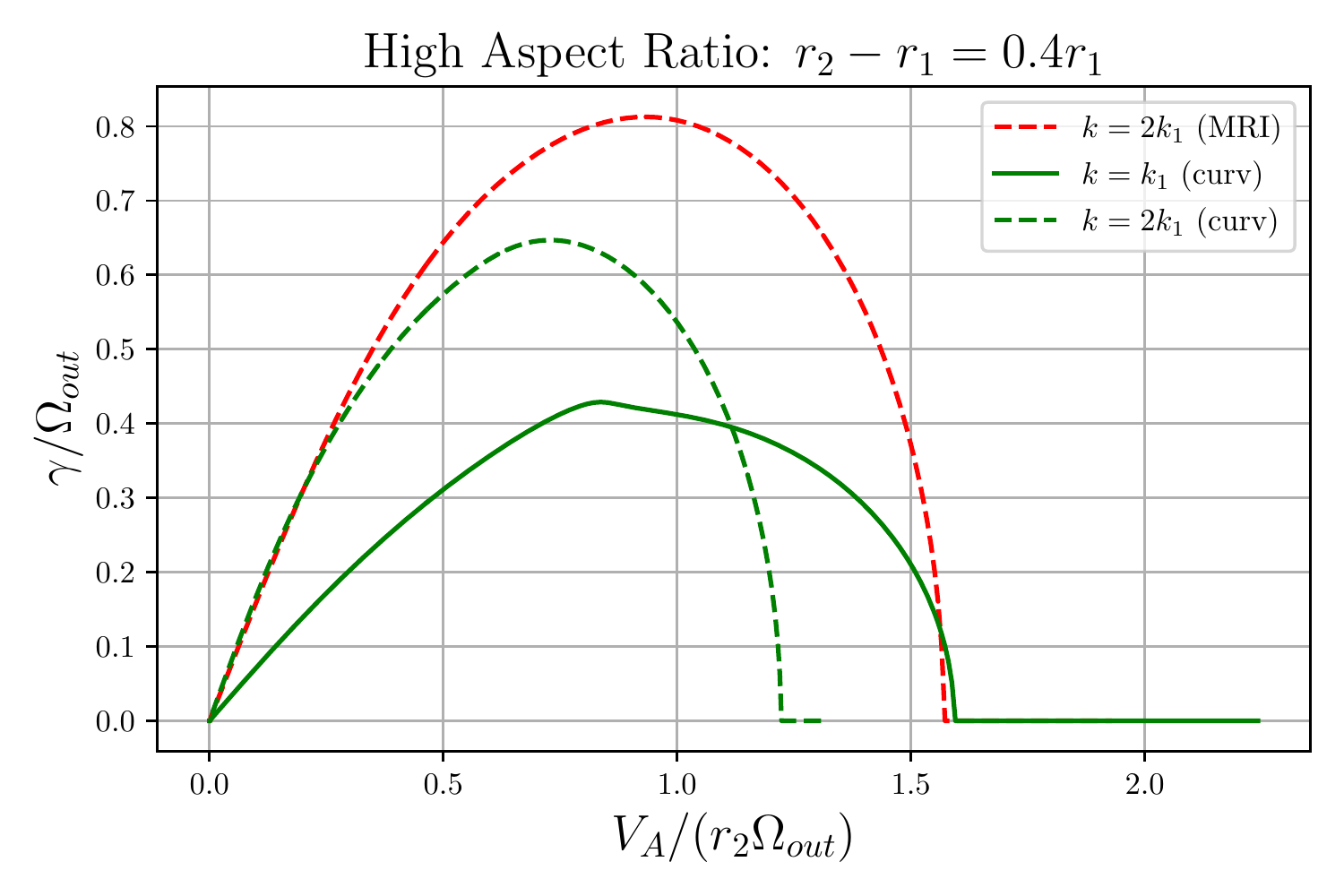}
        \centering
        (a)
        \includegraphics[width=\columnwidth]{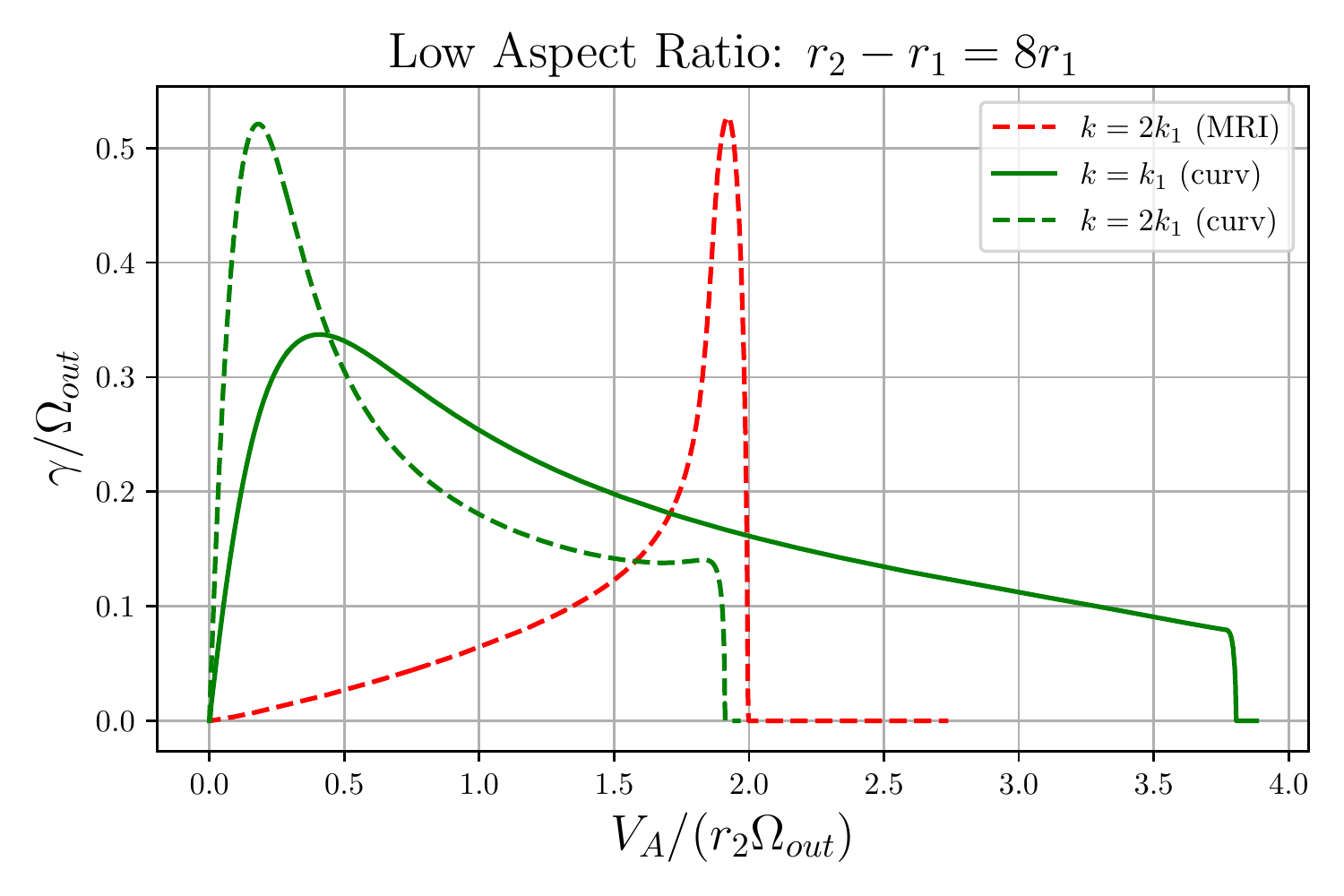}
        \centering
        (b)
        \includegraphics[width=\columnwidth]{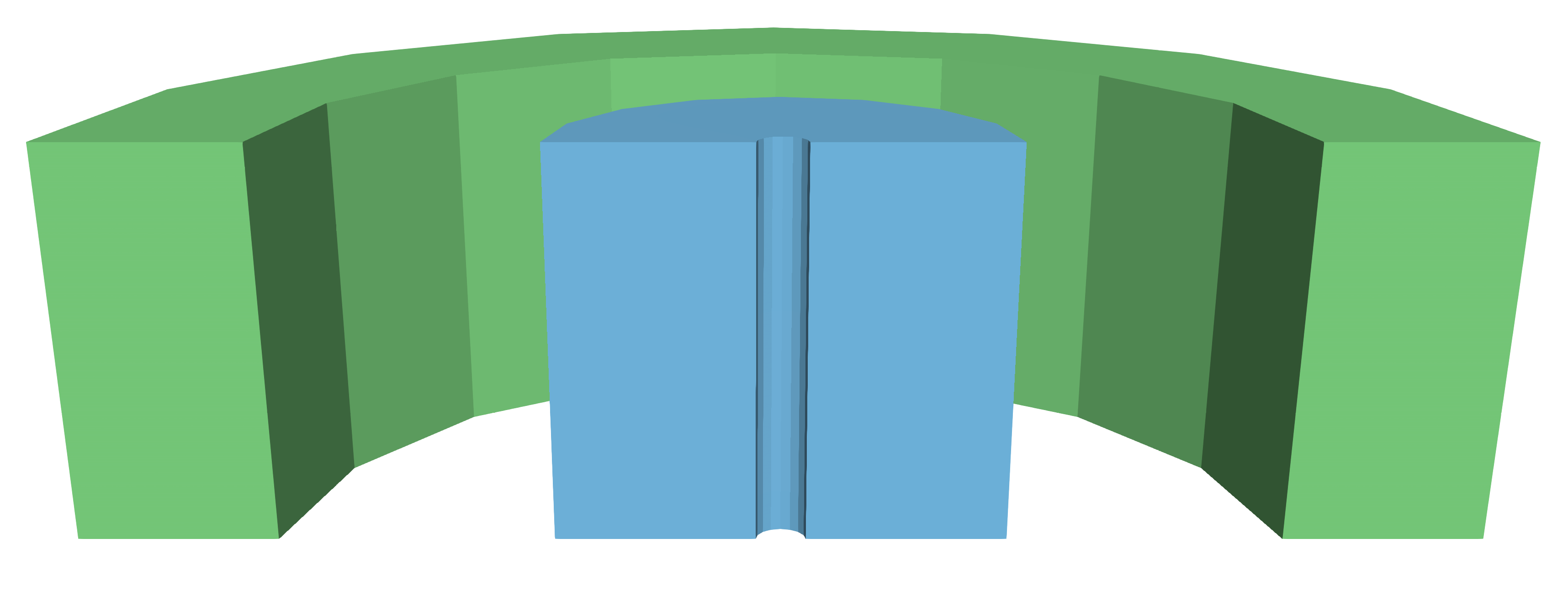}
        \centering
        (c)
    \caption{Growth rates in high and low aspect ratio $r_1/(r_2-r_1)$ with the curvature and MRI modes, showing the curvature mode's dominance in curved space and MRI's dominance in quasi-cartesian space. The normalization is against the \emph{outer} parameters here since these values change as a function of the average curvature. (c) shows how the aspect ratio affects the system geometry: the low and high aspect ratio cases are shown in blue and green respectively. Flow curvature and variation in shear is lower in the narrow gap (green) case and is therefore closer to a cartesian approximation.}
    \label{fig:gaplimit}
\end{figure}

The high aspect ratio case is seen in Fig.~\ref{fig:gaplimit} (a). It is apparent that as the system approaches quasi-cartesian geometry, the curvature mode is dominated by an everywhere-more-unstable MRI mode. The low aspect ratio case follows in Fig.~\ref{fig:gaplimit} (b), where it can be seen the range of magnetic fields where the MRI mode is dominant shrinks and the curvature mode dominates for most field values. This justifies our labeling of this mode as one caused by curvature effects. 

\subsection{Global solutions with pure azimuthal magnetic field}\label{sec:global_p}

In the presence of pure azimuthal magnetic field, the non-axisymmetric modes are the only accretion disk modes unstable ~\citep{terquem1996} to Keplerian shear flows. The global curvature effect of the system and magnetic field introduces additional radially dependent terms as shown in the WKB approximation in section \ref{sec:local}. Here, we numerically search for the global structures of non-axisymmetric modes through both the shooting method and linear simulations with the NIMROD code~\cite{SOVINEC2004355}. This would provide confidence in the results, as we would verify our solutions with both numerical techniques (shooting and NIMROD simulations). We start the simulations with a Keplerian flow and azimuthal magnetic field in both eigenvalue shooting and  initial-value NIMROD simulations.

\subsubsection{Solutions from global eigenvalue and linear simulations}
\begin{figure}
    \centering
        \includegraphics[width=\columnwidth]{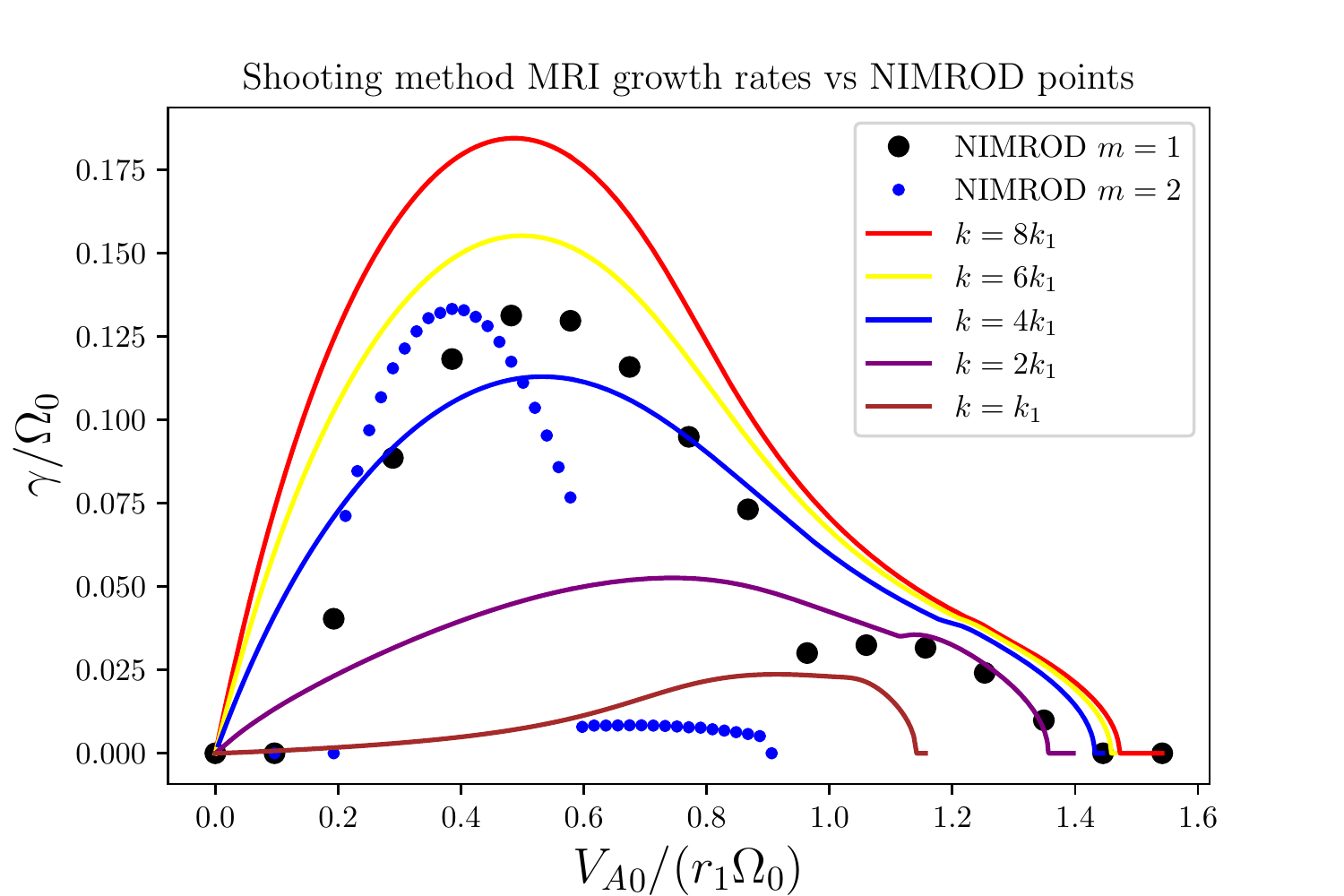}
    \centering
    (a)    
        \includegraphics[width=\columnwidth]{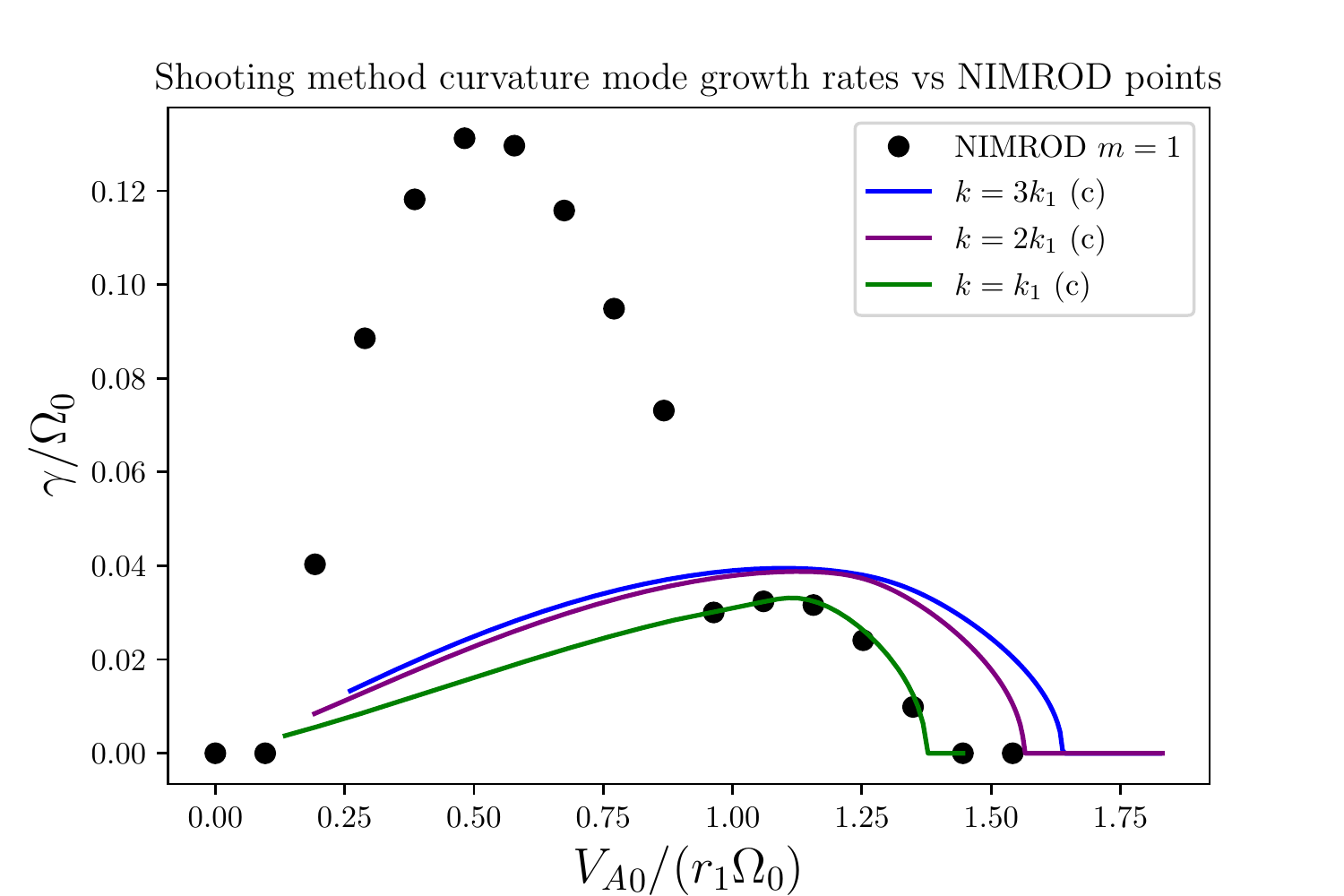}
    (b)
        \includegraphics[width=\columnwidth]{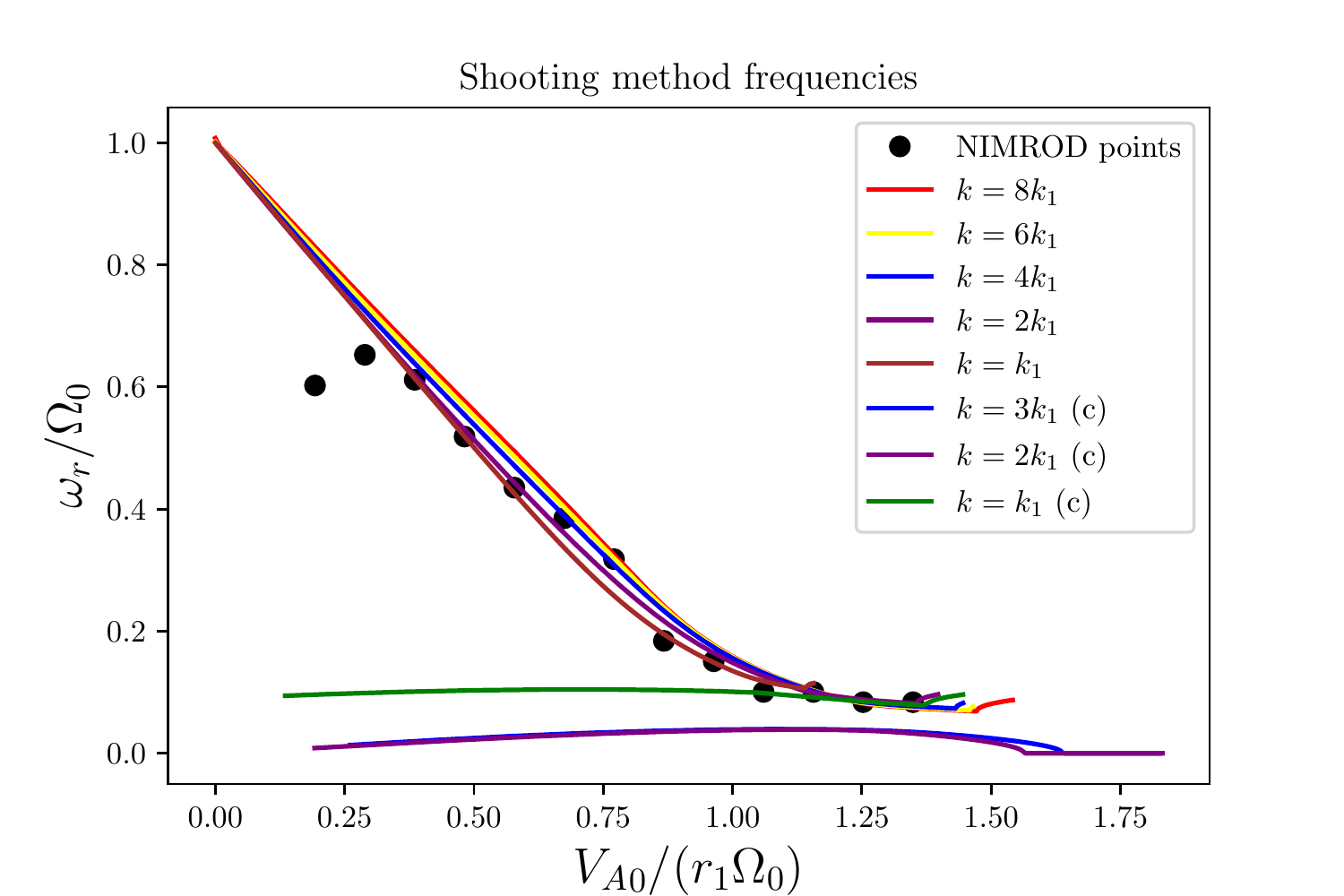}
    (c)
    \caption{(a), (b) Growth rates (c) frequencies from both global shooting method (eq. \ref{eq:ode}) and numerical linear NIMROD points for both typical MRI modes (with various $k_z$ in color) and global curvature modes (green curve). All shooting done for $m=1$.}
    \label{fig:bp_growthrates}
\end{figure}


\begin{figure}
    (a) MRI \hspace{35mm} (b) MRI\\
    \includegraphics[width=0.46\columnwidth]{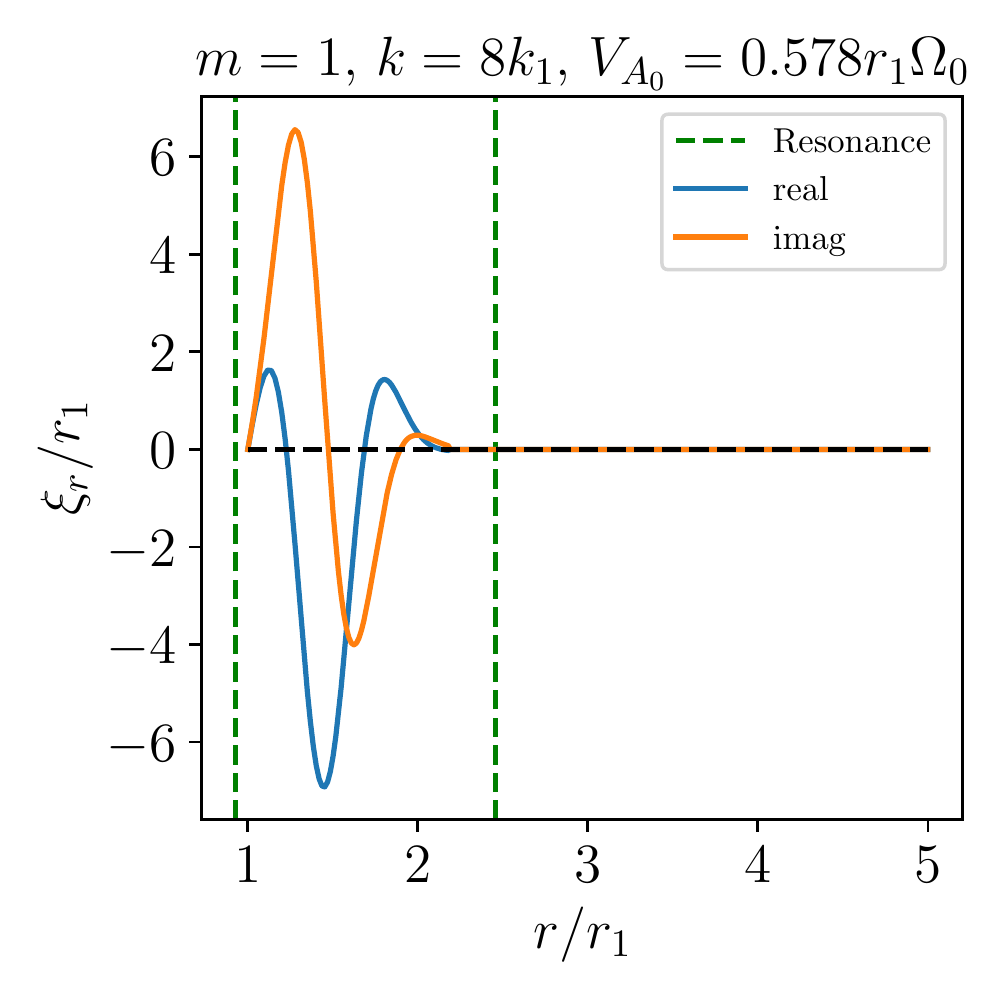} 
    \includegraphics[width=0.54\columnwidth]{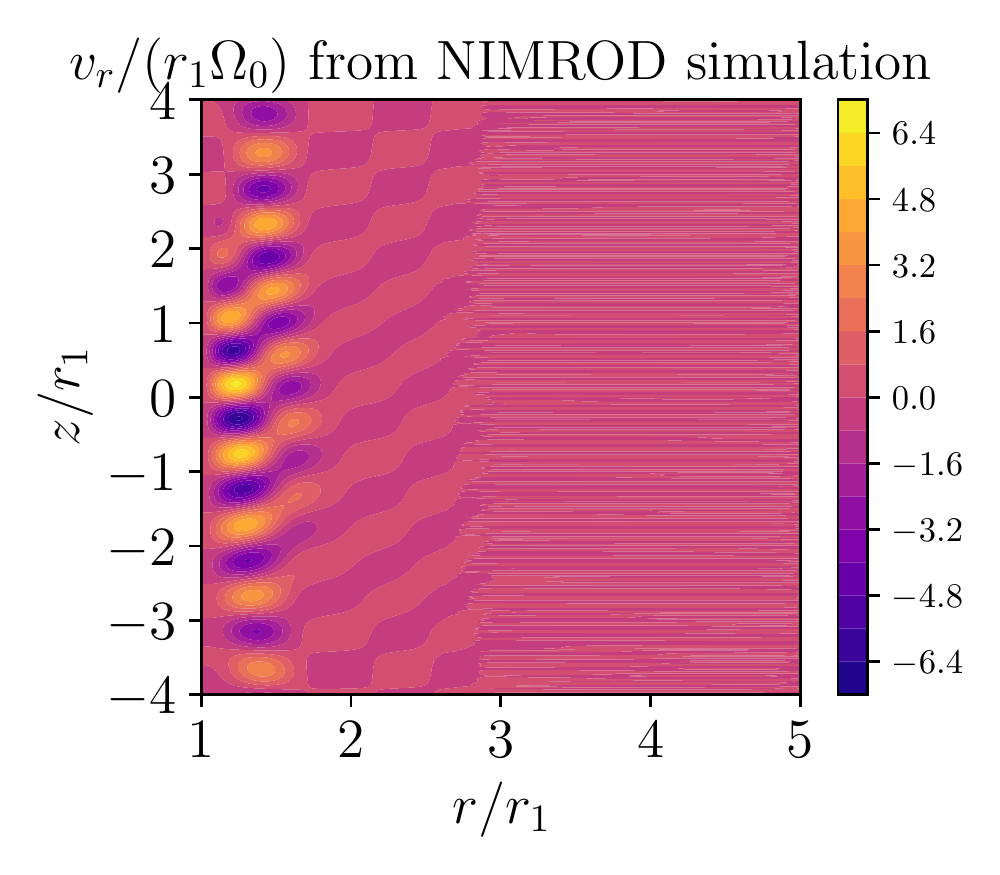}\\
    (c) curvature \hspace{28mm} (d) curvature\\
    \includegraphics[width=0.46\columnwidth]{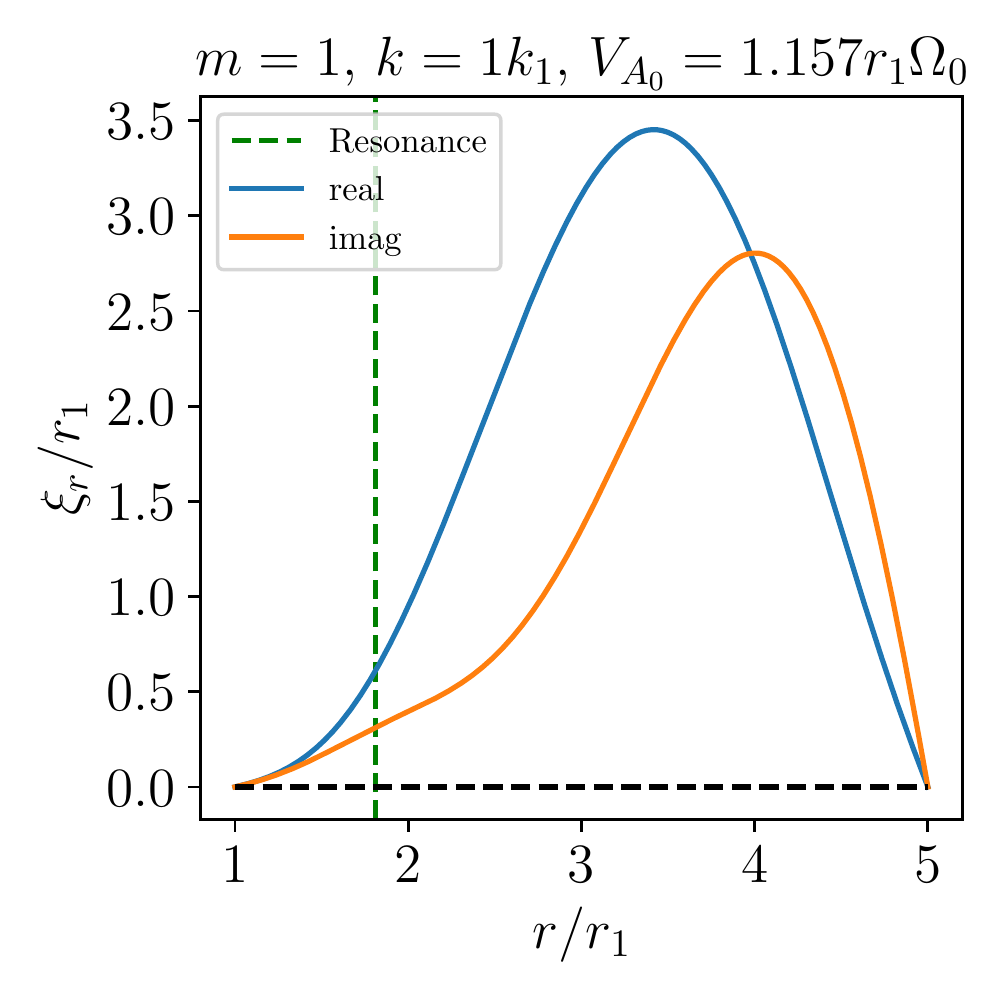}
    \includegraphics[width=0.54\columnwidth]{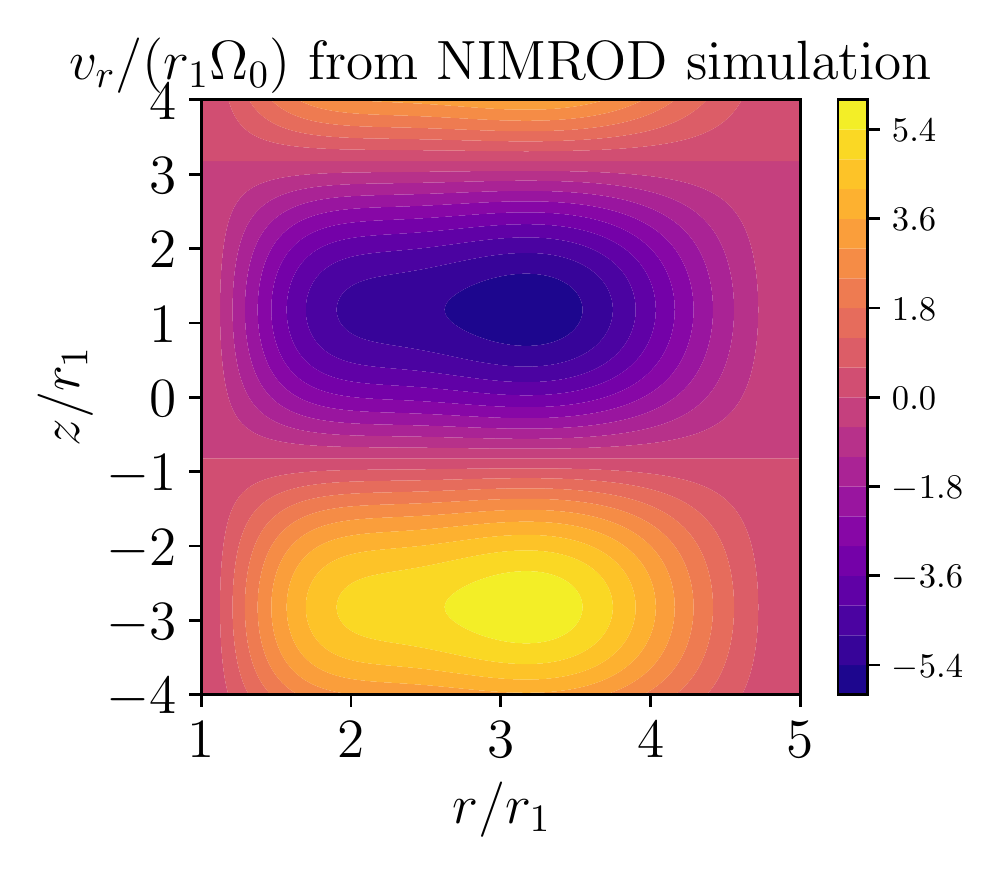}\\
      (e) mixed\\
     \vspace{-4mm}
    \includegraphics[width=0.75\columnwidth]{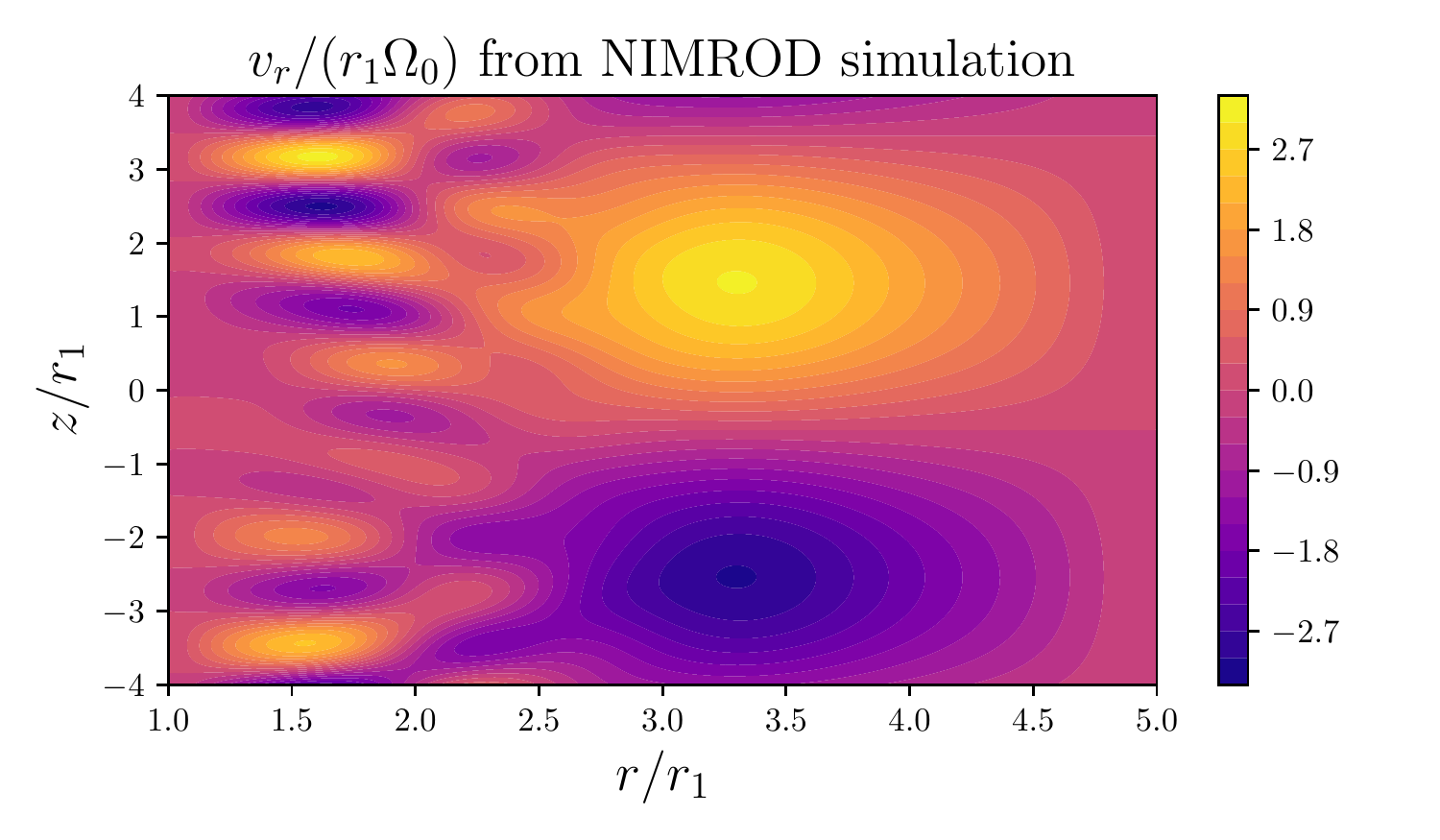}\\

    \caption{Linear $m=1$ mode structures in the toroidal case from eigenvalue shooting method [(a) and (c)] and equivalent NIMROD simulations [(b),(d),(f)]. (a) and (b)  the MRI mode at $V_{A_0} = 0.578 r_1 \Omega_0$;  (c) and (d) the curvature mode at $V_{A_0}= 1.157 r_1 \Omega_0$; and (e)  both modes mixed at $V_{A_0} = 0.868 r_1 \Omega_0$.}
    \label{fig:linear_modestructures}
\end{figure}

Figure~\ref{fig:bp_growthrates} shows growth rates and frequencies of the $m=1$ modes obtained from shooting and the NIMROD simulations. The solid curves show the results from shooting method and the numerical points are obtained by performing linear NIMROD simulations with the same initial state. By varying normalized Alfvén velocity (through changing $\B$) we observe that the growth rates of the MRI mode show a two hump curve similar to what was obtained via the dispersion relation shown in Fig.~\ref{fig:disp_rel_2d} (b). This is similar to what is shown in the growth rates of the NIMROD points. However, mode structures of the first 10 points (at low field $V_A/V_0<0.8$) resemble a system with $k=8k_1$ as shown in Fig.~\ref{fig:linear_modestructures} (b), which is consistent with the localized mode structure obtained from the shooting solutions at $k=8k_1$ (Fig.~\ref{fig:linear_modestructures} (a)). The eleventh point ($V_A/V_0\sim 0.86$) is transitional. Fig.~\ref{fig:linear_modestructures} (e) shows a local mode structure near the interior of the disk and a global mode structure at the edge of the disk, with each growing at similar magnitudes. This can be understood as the Alfvén velocity at which the MRI and curvature modes have growth rates of the same order. The points at Alfvén velocities above this show a dominant global mode, as seen in Fig.~\ref{fig:linear_modestructures} (c) and Fig.~\ref{fig:linear_modestructures} (d) for $V_A/V_0\sim 1.157$. This would indicate that for sufficiently high magnetic fields, the MRI mode is subject to higher dissipation in NIMROD simulations due to its local structure and higher vertical wavenumber (and loses more of its free energy at high fields from field line bending) and therefore the $k=k_1$ curvature mode dominates. 

Similar to the instability with vertical magnetic field in section \ref{sec:global_z}, here with only pure azimuthal B we also observe distinct global mode structure at high B. This is due to the curvature of the system (including magnetic field). The most global mode structure is obtained with the curvature mode as shown in Fig.~\ref{fig:linear_modestructures} (c). The equivalent contour plots of $m=1$ solutions obtained from NIMROD simulations are also shown in Fig.~\ref{fig:linear_modestructures}. Similarly, the transition from a local MRI (Fig.~\ref{fig:linear_modestructures} (b)) to a  global curvature mode (Fig.~\ref{fig:linear_modestructures} (d)) is observed. 
Interestingly, unlike eigenvalue shooting, NIMROD also captures the most unstable solutions in the 2D R-Z plane for a specific azimuthal number (here for $m=1$),  we observe a mixed solution of local MRI and global curvature at the transition point ($V_{A_0}=0.868$ $r_1 \Omega_0$) shown in Fig.~\ref{fig:linear_modestructures} (e). We therefore find that both the global shooting method and NIMROD solutions provide the same eigenvalues (growth rates and frequencies) as well as mode structures. A transition from more standard MRI mode to a very global mode structure (curvature mode) is obtained through both analyses. 

Here, we have not only verified the NIMROD solutions with the solutions obtained from Eq.\ref{eq:ode}, but both models have also produced new physics. This means that in many circumstances the curvature mode is newly found to be dominant and therefore contribute to momentum transport in differentially rotating systems. 

\begin{figure}
    \centering
        \includegraphics[width=0.46\columnwidth]{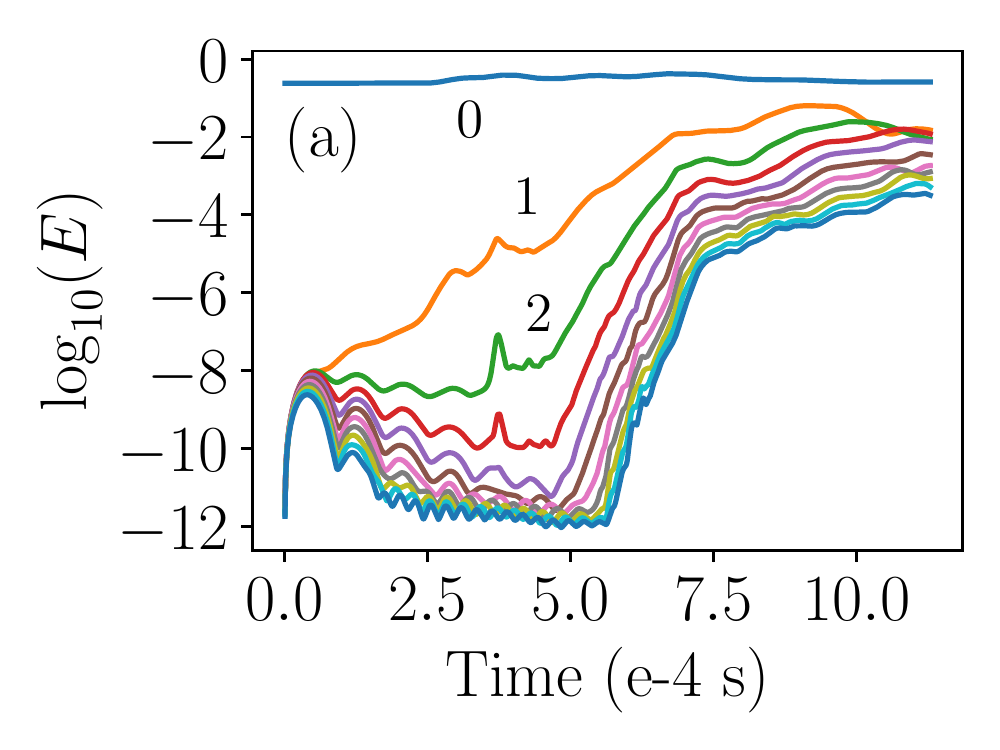}
        \includegraphics[width=0.46\columnwidth]{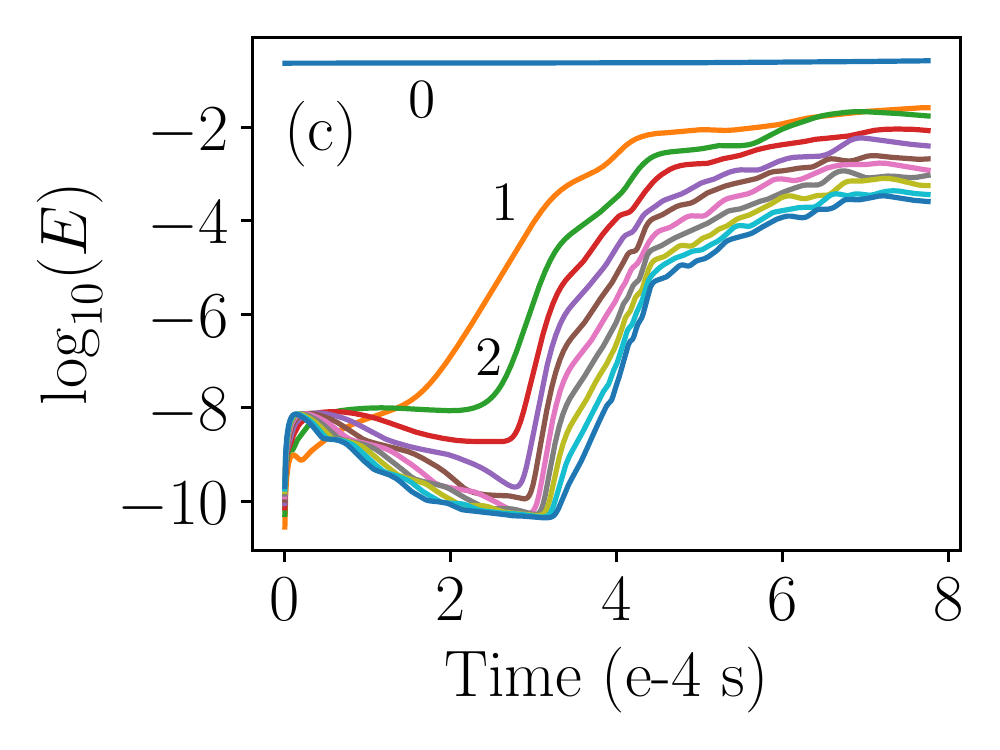}
        \includegraphics[width=0.46\columnwidth]{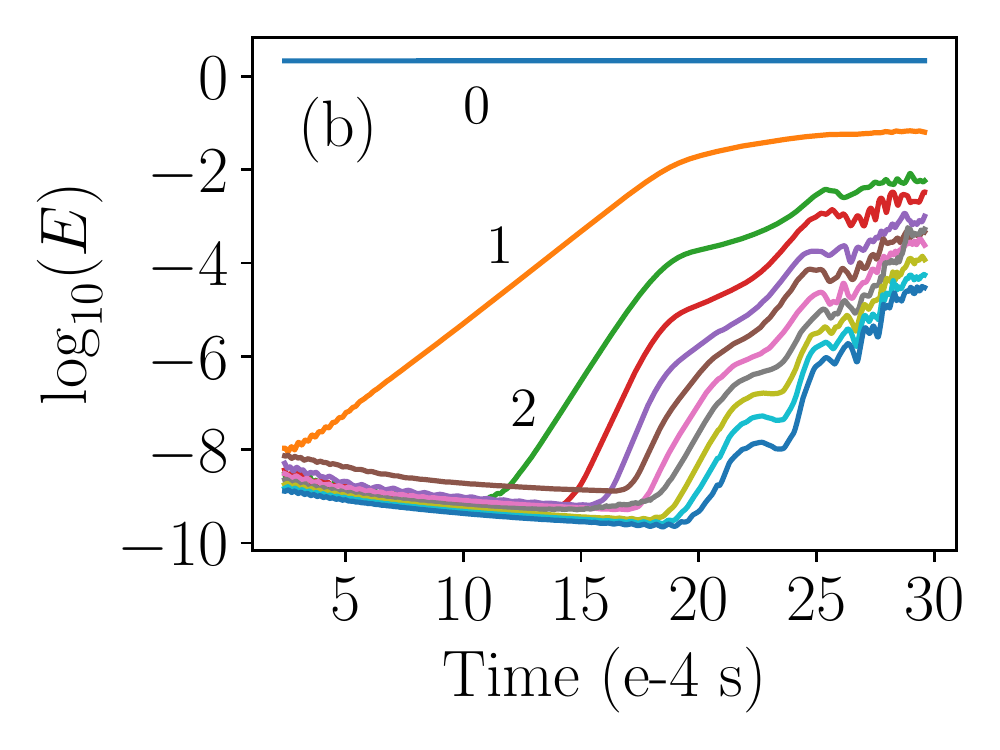}
        \includegraphics[width=0.46\columnwidth]{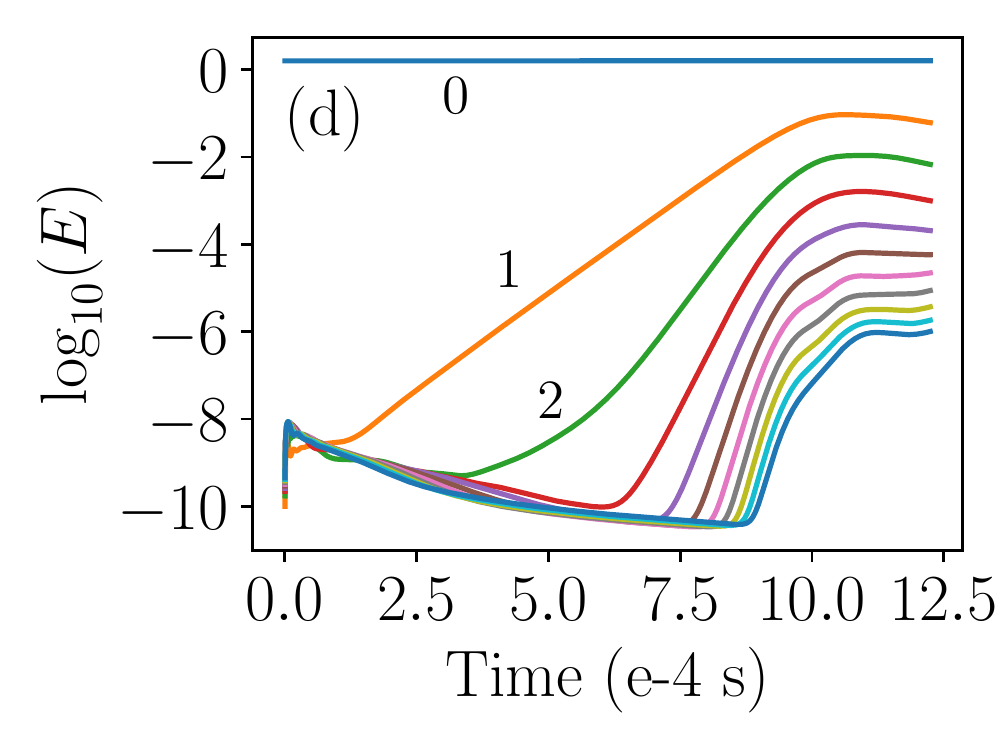}
        \includegraphics[width=0.46\columnwidth]{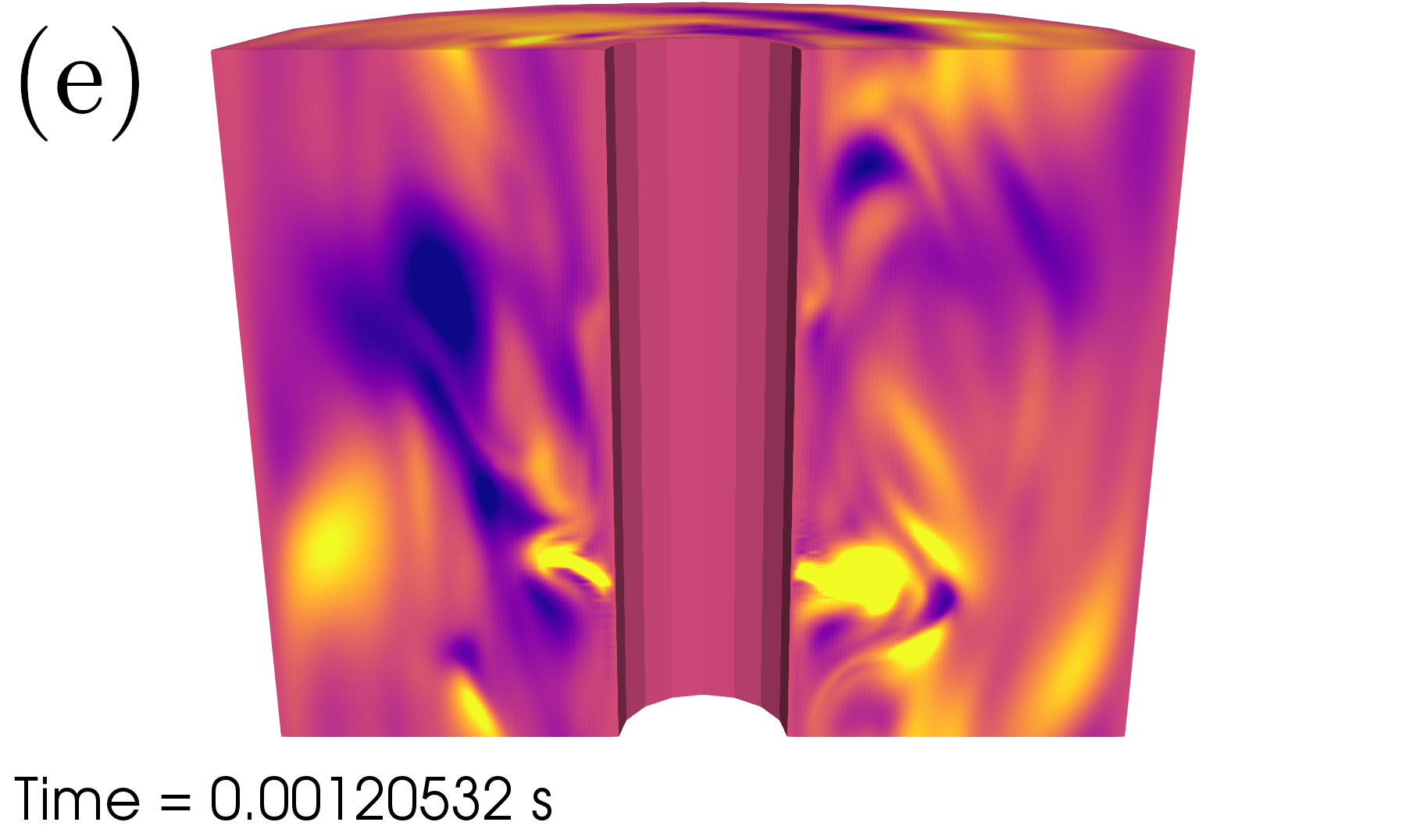}
            \includegraphics[width=0.46\columnwidth]{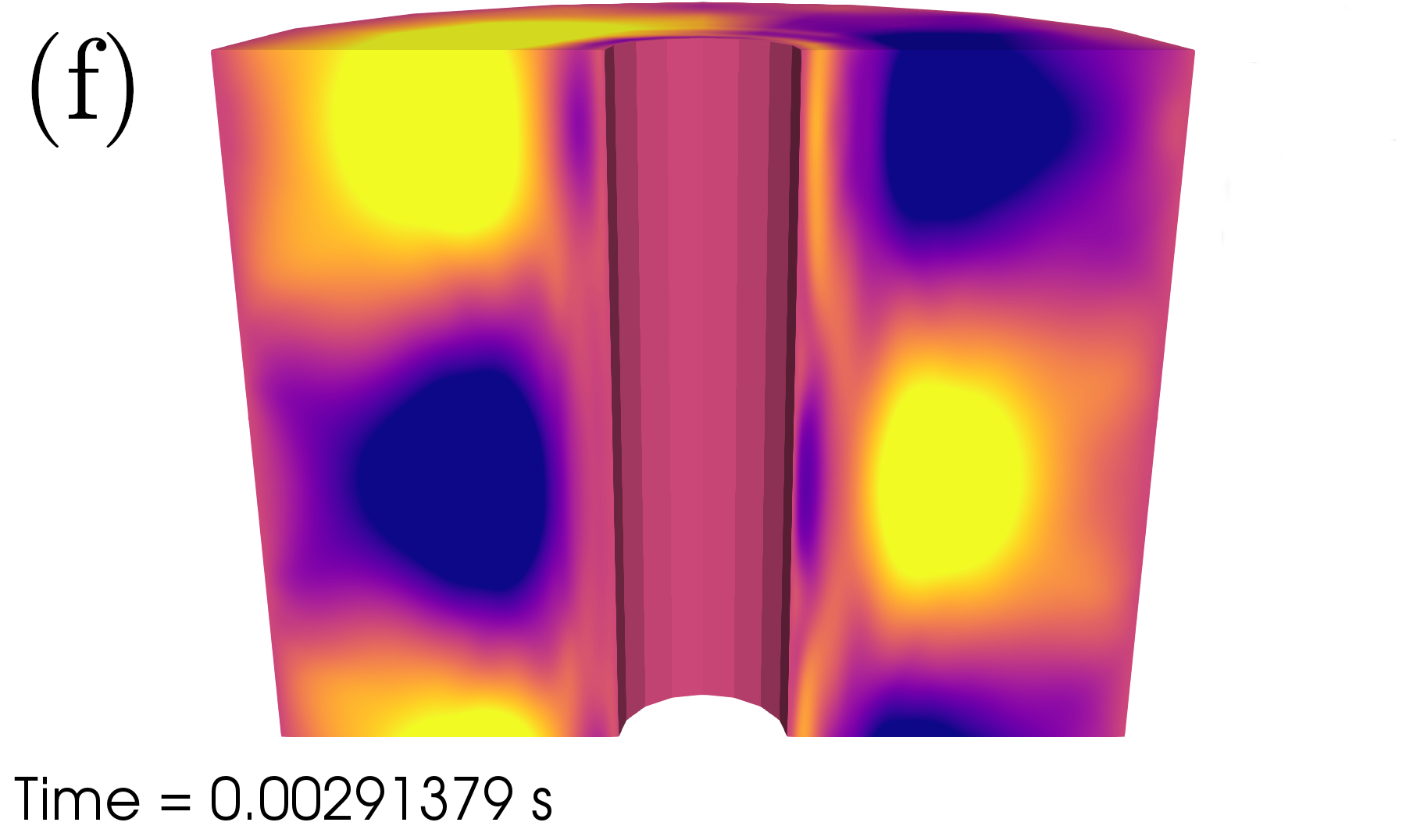}
          \begin{interactive}{animation}{PAF-merged.mp4}
        \includegraphics[width=0.46\columnwidth]{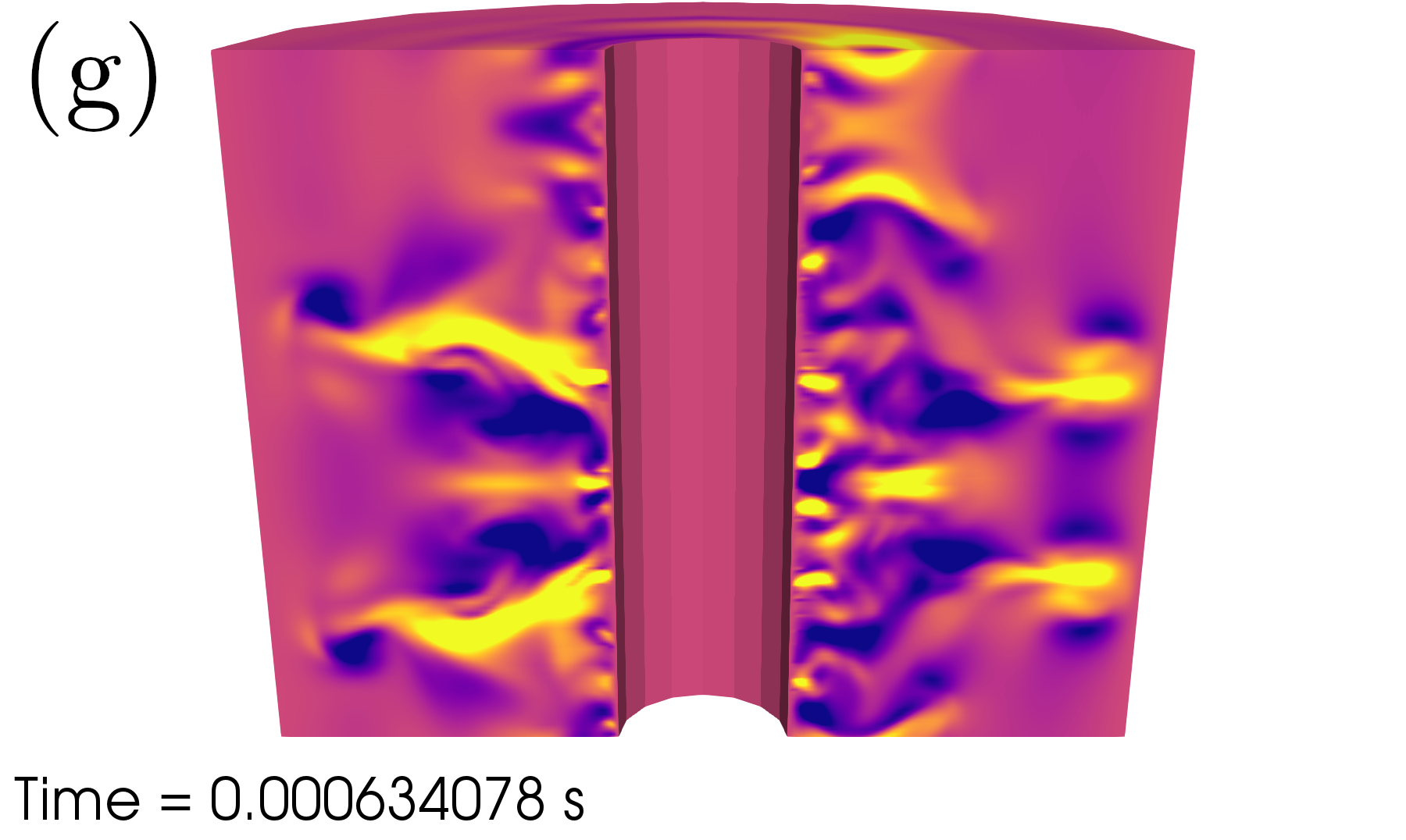}
        \includegraphics[width=0.46\columnwidth]{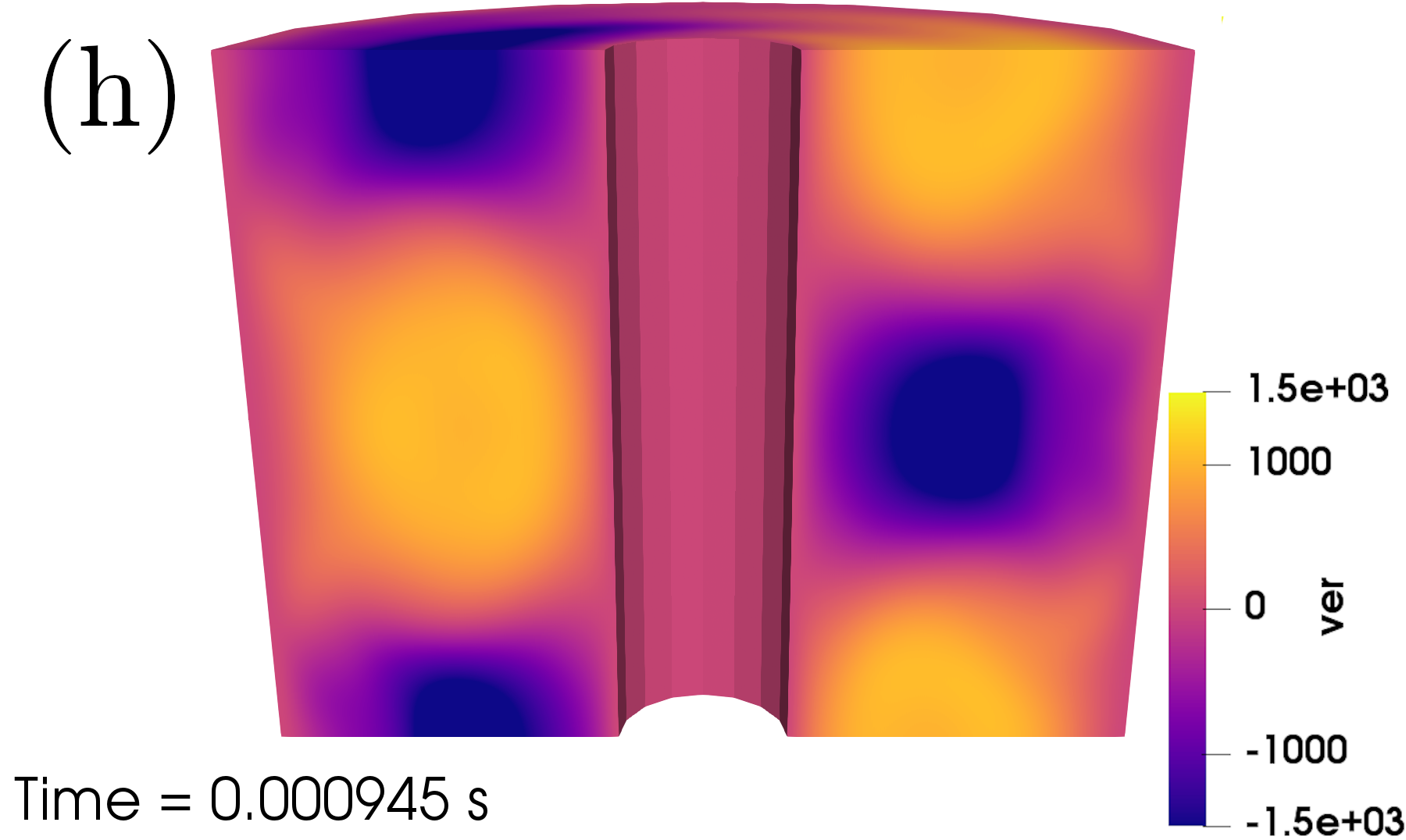}
            \end{interactive}
    \caption{Evolution of magnetic energy for several azimuthal modes (m=1,2,3 ...) from nonlinear simulations with   (a) low  ($V_A=0.2$ $r_1\Omega_0$) and (b) high ($V_A=0.58$ $r_1\Omega_0$) vertical field; and (c) low ($V_{A_0}=0.52$ $r_1\Omega_0$) and (d) high ($V_{A_0}=1.08$ $r_1\Omega_0$) toroidal field; nonlinear mode structures $v_r$ in a system with (e) low and (f) high vertical seed field, (g) low and (h) high toroidal seed field. An animated version of (g) and (h) for time-dependent solutions is available.}
    \label{fig:nonlin}
\end{figure}

\begin{deluxetable*}{cccccc}[tbh!]

    \tabletypesize{\footnotesize}

    
\tablehead{\colhead{Model} & \colhead{Rm=VL/$\eta$} & \colhead{Pm=$\nu$/$\eta$} & \colhead{S = $V_A L/\eta$} &\colhead{Resolution} & \colhead{L/R}}

\startdata 
Nonlinear global (VNF-MRI) & 12500 & 1 & 2500 & 40 $\times$ 40 deg.4 (43 modes)   &2\\
Nonlinear global (VNF-C) & 25000 & 1 & 14600 & 40 $\times$ 40 deg.4 (43 modes)   &2\\
Nonlinear global (PAF-MRI ) & 12500 & 1 & 6500 & 40 $\times$ 40 (86 modes) &2\\
Nonlinear global (PAF-C) & 15500 & 1 & 17000 & 80 $\times$ 80 deg. 4 (22 modes) & 2\\
Linear global & 25000 & 1 & 12000-20000 & 80 $\times$ 80 (160 $\times$ 160) deg 4 (m=1) & 1,2\\
\enddata
\caption{Parameters used in global NIMROD simulations. VNF= Vertical net flux, PAF= pure azimuthal (toroidal) field.}
\label{tab:simulations}
\end{deluxetable*}

\section{Full nonlinear simulations}\label{sec:nonlin}
To further investigate the transition from MRI-like modes to global curvature modes, we have performed 3D nonlinear MHD simulations. In contrast to the quasi-linear simulations where only one specific azimuthal mode number is linearly evolved (section \ref{sec:global}), here we evolve the global domain in the azimuthal direction by including many non-zero Fourier azimuthal modes (43 or 22 Fourier modes). Fig.~\ref{fig:nonlin} (a)-(d) show the evolution of  total volume-averaged magnetic energies for simulations started with vertical net flux (Fig.~\ref{fig:nonlin} a,b) and pure azimuthal field (Fig.~\ref{fig:nonlin} c,d). Consistent with the linear $m=1$ calculations (Fig.~\ref{fig:Bz_growthrates}), in the simulations started at low field where the localized MRI modes are dominant, we obtain several non-axisymmetric MRI modes grow and saturate to form a more turbulent state (Fig~\ref{fig:nonlin} a). A poloidal cross-section of turbulent structure during the saturation is shown in Fig~\ref{fig:nonlin} (e). This nonlinear simulation with weak vertical non-zero flux (VNF-MRI in table 1), with a dominant primary $m=0$ growth and the subsequent non-axisymmetric modes was extensively studied in \citep{rosenberg2021}. However, for simulations started at stronger magnetic field (at $V_A=0.58$ $r_1\Omega_0$, VNF-C in table \ref{tab:simulations} and shown with the star marker in Fig.~\ref{fig:Bz_growthrates}), the global $m=1$ curvature mode grows linearly first and saturates to form a more laminar state (Fig.~\ref{fig:nonlin} b). The cross-section of modal structure with a dominant global $m=1$ mode late during the saturation is shown in Fig.~\ref{fig:nonlin} (f). 

Similarly, for full nonlinear simulation with pure toroidal magnetic field, we obtain a transition from a turbulent to a laminar state as normalized magnetic field is increased. Fig.~\ref{fig:nonlin} (c)-(d) show the magnetic energy of several localized MRI axisymmetric modes growing to large amplitudes (Fig.~\ref{fig:nonlin} c), while for simulations at high field the magnetic energy is dominant by a global $m=1$ curvature mode (Fig.~\ref{fig:nonlin} d). Similarly, a transition from turbulent to a laminar saturate sate dominated with the $m=1$ curvature mode is also clearly seen through the cross-section of modal structures in Figs.~\ref{fig:nonlin} (g) and (h), respectively. The growth and the structure of the global $m=1$ mode in the nonlinear simulation is consistent with the linear curvature mode obtained with toroidal field in Fig.\ref{fig:linear_modestructures} (c,d).

\section{Summary and conclusions}\label{sec:summary}

Global stability of accretion flows in a differentially rotating Keplerian cylindrical disk is examined. Using a variety of linear local and global eigenvalue stability analyses, as well as linear and 3D nonlinear initial-value global simulations using the extended MHD NIMROD code, we uncover a global non-axisymmetric instability. Although this global mode co-exists with the local non-axisymmetric MRI modes, it persists at stronger magnetic fields as $V_A/V_0$ is increased, where local MRI is stable. As the mode structure and relative dominance over MRI of this mode is inherently determined by the global spatial curvature as well as the flow shear in the presence of magnetic field, we call it the magneto-spatial-curvature instability. 

Starting with the local WKB analysis of the non-axisymmetric modes in Keplerian accretion flows, we first observed the pronounced effect of the curvature terms due to the toroidal field curvature, or just spatial curvature terms in the vertical field case. Although WKB solutions provided some guidance, it clearly lacked the spatial radial variation of flow shear and field-line bending. We therefore moved to the global eigenvalue shooting method and linear NIMROD simulations to further investigate the effect of global curvature. Through global eigenvalue analysis (i.e. shooting solutions of the ODE (eq.~\ref{eq:ode}), we first found two distinct non-axisymmetric modes, a localized MRI concentrated in the high shear region and a new radially globally extended (with low m, and k) curvature mode.
The latter is absent in the local limit and persists at stronger magnetic field (with vertical or azimuthal fields). The  distinct nature of MRI and curvature modes is further demonstrated in the 2D plane of complex frequencies,  where MRI and curvature modes occupy different regions (curvature at lower frequency and MRI at high frequency), and co-exists for a wide range of $V_A/V_0$, until at larger values of $V_A/V_0$ (around unity), where  only curvature modes are present. Curvature modes are unstable Alfvén continua, which become global due to the inherent global shear flow and curvature.  The modified energy principle~\citep{frieman_rotenberg} was also examined, and we recovered the global differential rotation term (similar to $C_0$ in WKB) as the destabilizing effect. Additional terms due to the spatial curvature were also obtained. As we quasi-continuously change $m=1$ to $m=0$, we observe a bifurcation  of non-axisymmetric solutions (Fig.~\ref{fig:bifur}). We then inspected the modes' relative dominance in the low- and high-curvature regimes. By increasing the aspect ratio $r_1$/$\left(r_2-r_1\right)$, i.e. by moving the outer boundary closer in dimensionless quantities and therefore lowering the average spatial curvature, we approach a cartesian approximation and see the inner MRI mode dominate, whereas in the low-aspect-ratio limit, the global curvature mode dominates (Fig.~\ref{fig:gaplimit}).

Second, we also compared the global eigenvalue solutions with the direct linear simulations from NIMROD (i.e. the solution found by only evolving one azimuthal Fourier mode, $m=1$ for example), and found similar mode structures with the same growth rates (and real frequencies). In particular, NIMROD single-mode ($m=1$) growth rates with pure azimuthal field vs. $V_A/V_0$, also exhibit two humps, were the second hump occurs at stronger field due to curvature modes. 

Third, we finally performed full nonlinear simulations (where all Fourier modes m=0,1,2 ... 43 are evolved in the azimuthal direction) with the same initial condition as for the linear simulations. Consistent with the linear global solutions, a similar non-axisymmetric mode structure is found in the early phase of the nonlinear simulations. In addition, two distinct nonlinear saturated states are obtained. Full nonlinear simulations at stronger magnetic fields (as the ratio of $S/Rm = V_A/V_0$ approaches one and above), showed a global laminar nonlinear state, while a turbulent state is obtained in the simulations at weaker fields. 

In summary, in addition to the high-frequency and localized non-axisymmetric MRI modes (at high $m$ and $k_z$)~\citep{tajima05,Ogilvie_1996,goedbloed2022}, we find a previously  unidentified distinct global curvature mode (at low $m$ and $k_z$), which could be unstable at stronger field. Both local MRI and the curvature mode structures remain confined between the regions of Alfvénic points (where the magnitude of the Doppler-shifted wave frequency is equal to Alfvén frequency). However, due to the inherent presence of global flow shear and spatial curvature, the curvature mode extends throughout the domain.
The instability covers a wide range of parameter regions from weak to strong magnetic fields (i.e. from super-Alfv\'enic to sub-Alfv\'enic flows). Magneto-curvature instability is an Afvénic eigenmode instability, which is non-local due to global spatial and magnetic field curvature. The nonlinear dynamics in the presence of curvature modes (or the combination of local MRI and curvature modes) including the momentum transport and dynamo in accretion flows  will be topics of further investigations. In this paper, the cylindrical  geometry with real spatial curvature provided a rich family of unstable non-axisymmetric modes. The interplay between these modes and more classical local MRI modes is critical in understanding the process of transport of momentum, and the dynamo field generation ~\citep{brandenburg_1995,Rincon_2007,Ebrahimi_2009} and destruction of fields via magnetic reconnection ~\cite{rosenberg2021} in disks. Due to the less dissipative nature of the global curvature modes, our results could  therefore be relevant also to the experimental observations of different flow-driven modes in search of MRI in the laboratory~\citep{goodman02,wei2016,stefani2006,seilmayer2014,mishra2022,Yin-nature}, 
 and require further investigations.

\begin{acknowledgements}

Simulations were conducted on the cluster at the Princeton Plasma Physics Laboratory and the National Energy Research Scientific Computing Center (NERSC). This work was supported in part by funding from the Department of Energy for the Summer Undergraduate Laboratory Internship (SULI) program. This work is supported by the US DOE Contracts No. DE-AC02-09CH11466 and the Max-Planck-Princeton Center for Plasma Physics (MPPC).

\end{acknowledgements}
\bibliography{references}
\bibliographystyle{aasjournal}

\end{document}